\documentclass[preprint,prd,showpacs]{revtex4}
\usepackage{amsmath,epsfig}
\usepackage{amsfonts}
\usepackage{amssymb}
\usepackage{graphicx}
\usepackage[cmtip,arrow]{xy}

\usepackage{amssymb,amsopn}

\begin{document}
\title{Deformation quantization of cosmological models}

\author{Rub\'en Cordero}\email{cordero@esfm.ipn.mx}
\affiliation{Departamento de F\'{\i}sica, Escuela Superior de
F\'{\i}sica y Matem\'aticas del IPN\\ Unidad Adolfo
L\'opez Mateos, Edificio 9, 07738, M\'exico D.F., M\'exico}

\author{Hugo Garc\'{\i}a-Compe\'an}
\email{compean@fis.cinvestav.mx} \affiliation{Departamento
de F\'{\i}sica, Centro de Investigaci\'on y de Estudios
Avanzados del IPN\\ P.O. Box 14-740, 07000 M\'exico D.F., M\'exico}

\author{Francisco J. Turrubiates}\email{fturrub@esfm.ipn.mx}
\affiliation{Departamento de F\'{\i}sica, Escuela Superior de F\'{\i}sica
y Matem\'aticas del IPN\\ Unidad Adolfo
L\'opez Mateos, Edificio 9, 07738, M\'exico D.F., M\'exico.}
\date{\today}

\begin{abstract}
The Weyl-Wigner-Groenewold-Moyal formalism of deformation
quantization is applied to cosmological models in the minisuperspace. The
quantization procedure is performed explicitly for quantum cosmology
in a flat minisuperspace. The de Sitter cosmological model is worked out in detail and
the computation of the Wigner functions for the Hartle-Hawking,
Vilenkin and Linde wave functions are done numerically. The Wigner function is analytically calculated for the Kantowski-Sachs model in (non)commutative quantum cosmology and for string cosmology with dilaton exponential potential.  Finally, baby universes solutions are described in this context and the Wigner function is obtained.

\end{abstract}
\vskip -1truecm
 \maketitle

\vskip -1.3truecm
\newpage

\setcounter{equation}{0}

\section{Introduction}

One of the fundamental problems of cosmology is the initial
singularity. It is commonly believed that near this singularity the physical
evolution is governed by quantum mechanics. In the quantum cosmology
framework the whole universe is represented through a wave function
satisfying the Wheeler-DeWitt equation \cite{DeWitt}. It is well
known that it represents the low energy approximation of string
theory, however it contains some non-trivial leading order
information.  The development of quantum cosmology started at the
beginning of the 80's of last century and one of its main ideas is
that the universe could be spontaneously nucleate out from nothing
\cite{Tryon,Fomin,Atkatz,Alex1,Zeldovich,HawkingT,LindeT,Rubakov,AlexT}.
After nucleation the universe can enter into a phase of inflationary
expansion and continues its evolution to the present time. However,
there are several important questions that remain to be solved like
the general definition of probability, time and boundary
conditions \cite{approch}. In order to find a unique solution of the
Wheeler-DeWitt equation it is necessary to impose
boundary conditions. In the case of quantum mechanics there is an
external setup and the boundary conditions can be imposed safely,
however in 4-dimensional quantum cosmology there is nothing external
to the  universe and the question of which one is the correct
prescription for the boundary condition of the universe is controversial \cite{debate}. There are several proposals for the correct boundary conditions in quantum cosmology, for example, the no-boundary proposal of Hartle and Hawking \cite{HawkingT}, the tunneling proposal of Vilenkin \cite{AlexT} and the proposal of Linde \cite{LindeT}. In a recent development in quantum cosmology a principle of selection in the landscape of string vacua has been proposed in  \cite{Brustein:2005yn}, in the context of the minisuperspace. Moreover in \cite{Kounnas:2007pg}
different cosmologies were defined in terms of a wave function on a compact worldsheet of system of parafermions. Much of the information is encoded in the wave function as function of the moduli space.

Another important issue is how to extract information of the Wheeler-DeWitt equation. In general, the configuration space used in quantum cosmology is an infinite dimensional space called superspace and it is not amenable to work with. In the study of homogeneous universes the infinite dimensional space is truncated to finite degrees of freedom, therefore obtaining a particular minisuperspace model. The reduction of superspace to minisuperspace is not a rigorous approximation scheme, however there is the hope that minisuperspace maintains some of the essential features of quantum cosmology. In classical cosmology homogeneity and isotropy are fundamental to describe the universe at large scale and therefore it is expected to have a minisuperspace description in quantum cosmology.

At very early times of the universe, even before the Planck time, the universe should be described by means of quantum cosmology that could take into account effects from string theory, supersymmetry and noncommutativity. String cosmology \cite{Gasperini}, quantum wormholes, baby universes and supersymmetric quantum cosmology has become very intensive research areas in quantum cosmology. Quantum wormholes \cite{Strominger} are instanton solutions and are important in the Euclidean path integral formulation. In the third quantization approach, which is an adequate description of topology change in the path integral quantization, the wave function is transformed into a quantum field operator which includes operators that create and destroy the so called baby universes.

Supersymmetric quantum cosmology is one of the main research areas \cite{Macias:1993mq,Graham,Obregon:1996dt,D'Eath:1996at}. At the time of quantum creation of the universe it is possible that supersymmetry would not yet be broken and therefore could have very important consequences in the evolution of the universe. Other very interesting effects that could arise in the very early universe are the effects of noncommutativity \cite{GarciaCompean:2001wy}.

At the end of the late 1980s and early 1990s quantum decoherence, the transition from quantum physics to classical physics, was an active research area in quantum cosmology \cite{HartleBU,HalliwellBU} (for more recent reviews see, \cite{Halliwell:2005xj,thorwart}). In  Refs. \cite{Halliwell,Singh-Padma,Kodama,Calzetta:1989vk,Habib:1990hx}, quantum cosmology is developed in the phase space and the use of Wigner function is showed to be a very useful approach to study decoherence. In fact, quantum mechanics in phase space is an appropriate formalism to describe
quantum mechanical systems (for a review see \cite{hillery}). The description using the Wigner function has been of considerable interest and usefulness in quantum optics, nuclear and particle physics, condensed matter, statistical physics, etc. In particular, the description of semiclassical properties and the analysis of the classical limit is more clear in the Wigner function formalism.
The classical limit from quantum cosmology was studied also in \cite{Habib:1990hx,Habib:1990hz}, where the use of Wigner function and quantum mechanics in phase space is fundamental. It has been argued that quantum decoherence is achieved if the density matrix is coarse grained, i.e. averaged over configuration or phase space variables, or in an alternative way, if the system interact with some environment which is not monitored. However the existence of classical correlations is another characteristic of the classical limit and requires the presence of sharp peaks of the Wigner function, but a coarse graining produces a spreading of the distribution in phase space. The former arguments demand a subtle coarse graining in which there is a delicate balance between the existence of classical correlations and decoherence. A coarse graining could be modeled by a Liouville equation with friction and diffusion terms \cite{Habib:1990hx}. Some other further results are described in  Refs. \cite{LaFlamme:1990kd,Halliwell:1992gk}.

The quantum mechanics in phase space is only one part of a complete and consistent type of quantization termed: {\it deformation quantization}. In this paper we formulate quantum cosmology in terms of deformation quantization and rewrite some results of \cite{Habib:1990hx,Habib:1990hz} in the context of this formalism. We will assume that the superspace (of 3-metrics) is flat and a Fourier transform can be defined. This is fairly valid for the ample set of examples in two dimensions that we present in this paper. The case of curved minisuperspace will be discussed in a future communication.

The deformation quantization formalism is an alternative approach to
the canonical and path integral quantizations and has its origins in
the seminal works by Weyl \cite{weyl}, Wigner \cite{wigner},
Groenewold \cite{groene},  Moyal \cite{moyal}, and Vey \cite{vey},
and is based on the idea of treating with quantum mechanics on the
phase space. In 1978 Bayen, Flato, Fronsdal, Lichnerowicz and
Sternheimer \cite{bayen} introduced its final form in which the
quantization is understood as a deformation of the classical algebra
of observables instead of a change in the nature of them. This
quantization arises as a deformation of the usual product algebra of
the smooth functions on the classical phase space and then as a
deformation of the Poisson bracket algebra. The deformed product is
called the $\star -$product which has been proved to exist for any
symplectic manifold \cite{prueba}, \cite{fedosov} and more recently
shown by Kontsevich \cite{k} that it also exists for any Poisson
manifold. These results in principle allow us to perform the
quantization of arbitrary Poissonian or symplectic systems and to
obtain in an easier way the classical limit due to the nature of the $\star
-$product. Our aim is to introduce this formalism in quantum
cosmology and apply the results to simple models.

The structure of the paper is as follows. In section II we survey
the canonical Hamiltonian formalism of general relativity and its
canonical quantization. In section III we construct the
Stratonovich-Weyl quantizer, the star-product and the Wigner
functional. Section IV is devoted to apply the deformation
quantization procedure to several minisuperspace models and to obtain their Wigner function. In subsection IV.A we treat de Sitter model and calculate numerically the Wigner function for the Hartle-Hawking, Vilenkin and Linde wave functions. In subsection IV.B we calculate analytically the Wigner function for the Kantowski-Sachs model in (non)commutative quantum cosmology and we interpret the results. In subsection IV.C, we obtain the Wigner function
for string cosmology with dilaton exponential potential in terms of the Meijer's function. Besides, in subsection IV.D the Wigner function for the baby universes solutions is calculated by means of the Moyal product and
the annihilation and creation operators. Finally, in section V we give our final remarks.

\section{Canonical Formalism of General Relativity}

In this section we briefly overview the Hamiltonian formalism of
general relativity. Our presentation will not be complete and
detailed information can be found in Ref. \cite{DeWitt,IntQC}. Our intention is
only to introduce the notation and conventions for future reference
along the paper.

\subsection{Hamiltonian Formalism}

We start from the ADM decomposition of general relativity and consider
a pseudo Riemannian manifold $(M,g)$ which is globally hyperbolic.
Thus, spacetime $M$ can be decomposed as $M= \Sigma \times
\mathbb{R}$ where $\Sigma$ is an spatial hypersurface and the
metric of a foliation is: $ds^2=g_{\mu \nu} dx^\mu
dx^\nu= - (N^2-N^iN_i) dt^2 + 2N_idx^idt + h_{ij} dx^i dx^j,$ with
signature $(-,+,+,+).$ Here $h_{ij}$ is the intrinsic metric on the
hypersurface $\Sigma$, $N$ is the lapse function and $N^{i}$ is the shift vector.

The space of all Riemannian 3-metrics and scalar matter configurations $\Phi$ on
$\Sigma$ is the so called {\it superspace} ${\tt Riem}(\Sigma)=\{
h_{ij}(x), \  \Phi(x) | x \in \Sigma \}.$ Let us denote the space of
Riemannian metrics on $\Sigma$ as ${\tt Met}(\Sigma)=\{ h_{ij}(x) |
x \in \Sigma\}$ which is an infinite dimensional manifold. The
moduli space ${\cal M}$, is then defined as the configuration space
(superspace) modulo the group of diffeomorphisms ${\tt
Diff}(\Sigma)$ i.e. ${\cal M}={\frac{{\tt Riem}(\Sigma)}{{\tt
Diff}(\Sigma)}}$ or for pure gravity ${\cal M}={\frac{{\tt
Met}(\Sigma)}{{\tt Diff}(\Sigma)}}$.

The corresponding phase space (Wheeler's phase superspace) $\Gamma^*\cong T^* {\tt Met}(\Sigma)$
is given by the pairs $\Gamma^*= \{(h_{ij}(x), \pi^{ij}(x))\}$,
where $\pi^{ij}=\frac{\partial{L}_{EH}}{\partial \dot{h}_{ij}}$ and
${L}_{EH}$ is the Einstein-Hilbert Lagrangian. In the following we
will deal with fields at the moment $t=0$ (on $\Sigma$) and we put
$h_{ij}(x,0) \equiv h_{ij}(x)$ and $\pi^{ij}(x,0) \equiv
\pi^{ij}(x)$.

The dynamics of general relativity coupled to matter in this
foliation is encoded in the variation of the following action
\begin{equation}
{\cal S}= \int dt L =   \frac{1}{16 \pi G_N} \bigg[ \int_M d^4x \sqrt{-g} \
\big( {^4R}(g) - 2 \Lambda \big) + 2 \int_{\partial M} d^3x \sqrt{h}
K \bigg] + {\cal S}_m, \label{GRAction}
\end{equation}
where ${^4R}(g)$ is the scalar curvature in four dimensions
depending on the spacetime pseudo-Riemannian metric $g_{\mu\nu}$, $g$ is the determinant of $g_{\mu\nu}$, $G_{N}$ is the gravitational constant in $N$ dimensions, $\Lambda$
is the cosmological constant, $K$ is the trace of the extrinsic curvature $K^i_i$ compatible with $h_{ij}$
and ${\cal S}_m$ is the matter action. For the case of scalar field matter
subject to a potential $V(\Phi)$ we have that
\begin{equation}
{\cal S}_m = \int_M d^4 x \sqrt{-g} \bigg(-{\frac{1}{2}} g^{\mu \nu}
\partial_{\mu}\Phi \partial_\nu \Phi - V(\Phi) \bigg).
\end{equation}
The action (\ref{GRAction}) can then be written as
\begin{equation}
{\cal S}= \int dt d^3x\bigg( \pi^0 \dot{N} + \pi^{i} \dot{N}_i
-N {\cal H}_\perp - N^i {\cal H}_i \bigg),
\end{equation}
where ${\cal H}_\perp$ and ${\cal H}_i$ are given below.
Thus the canonical Hamiltonian is given by
$$
H_C = \int d^3x\bigg( \pi^0 \dot{N} + \pi^{i} \dot{N}_i + \pi^{ij}
\dot{h}_{ij} + \pi_{\Phi} \dot{\Phi} \bigg) - L
$$
\begin{equation}
= \int d^3x\bigg( \pi^0 \dot{N} + \pi^i \dot{N}_i + N {\cal
H}_\perp +N^i {\cal H}_i \bigg),
\end{equation}
where $\pi^{0}=\frac{\partial L}{\partial \dot{N} }$ and $\pi^{i}=\frac{\partial L}{\partial \dot{N^{i}} }$ and $\pi_{\Phi}=\frac{\partial L}{\partial \dot{\Phi}}$.
The corresponding equations of motion for $N$ and $N_i$ yield the Hamiltonian constraint and the momentum constraint, respectively
\begin{eqnarray}
\hspace*{-3cm}{\cal H}_\perp(x) &=& {h}^{-1/2} \bigg(\frac{1}{2} \pi^2 - \pi^i_j \pi^j_i \bigg)
+ \sqrt{h} \ ^3R \,\, \nonumber \\
&=& 4 \kappa^2 G_{ijkl}\pi^{ij} \pi^{kl} - \frac{\sqrt{h}}{4 \kappa^2} \big({^3R} - 2 \Lambda\big) + \frac{1}{2} \sqrt{h}
\bigg(\frac{\pi^2}{h} + h^{ij} \Phi_{,i} \Phi_{,j} + 2 V(\Phi) \bigg) =0
\label{eqn:constraints1}\,\,, \\
{\cal H}_i(x) &=& \frac{\sqrt{h}}{2\kappa^{2}}\left(G^{0} _i -2\kappa^{2}T^{0} _i\right)= -2 \pi^{ij}_{|j} + h^{ij} \Phi_{,j} \pi_{\Phi} = 0,
\label{eqn:constraints2}
\end{eqnarray}
where $\kappa^{2} = 4\pi G_{N}$, ${^3R}(h)$ is the scalar curvature of $\Sigma$, $G_{ijkl} =
\frac{1}{2}h^{-1/2} \big(h_{ik}h_{jl} + h_{il} h_{jk} - h_{ij}
h_{kl} \big)$, ${}_{|j}$ denotes covariant derivative with
respect to $h_{ij}$ and $h$ is its determinant.

In this way the Poisson bracket between $h_{ij}$ and $\pi^{kl}$ is given by
\begin{equation}
\{h_{ij}(x), \pi^{kl}(y) \}_{PB}= \frac{1}{2} (\delta^k_i \delta^l_j
+ \delta^k_j \delta^l_i) \delta(x-y)\,\, ,
\end{equation}
which is one of the most important structures for quantization.

\subsection{Canonical Quantization}

The most employed formalism to quantize a physical system is the canonical quantization which can be applied in general relativity. In this section we describe briefly the general procedure to obtain the quantum equations.  In the $h$-representation, the usual promotion of canonical coordinates $h_{ij}(x),$ $\Phi(x)$ and $\pi^{ij}$,
$\pi_{\Phi}$ to the operators can be done
 in the following form: $\widehat{h}_{ij} |h_{ij},\Phi \rangle =h_{ij} |h_{ij},\Phi \rangle$,
$\widehat{\pi}^{ij} |h_{ij},\Phi \rangle =-i \hbar
\frac{\delta}{\delta h_{ij}(x)} |h_{ij},\Phi \rangle,$
$\widehat{\Phi} |h_{ij}, \Phi \rangle =\Phi(x) |h_{ij},\Phi \rangle$
and $\widehat{\pi}_\Phi |h_{ij},\Phi \rangle =-i \hbar
\frac{{\delta}}{\delta \Phi(x)} |h_{ij},\Phi \rangle.$ These
operators satisfy the commutation relations

\begin{equation}
[\widehat{h}_{ij}(x), \widehat{\pi}^{kl}(y)] = \frac{i \hbar}{2} (\delta^k_i \delta^l_j + \delta^k_j \delta^l_i)
\delta(x-y).
\end{equation}
The constraints (\ref{eqn:constraints1}) and (\ref{eqn:constraints2}) have to be imposed at the quantum level in the form
\begin{equation}
\widehat{\cal H}_\perp | \Psi \rangle =0, \ \ \ \ \ \ \ \ \ \ \ \
\widehat{\cal H}_i| \Psi \rangle =0 \,\, .
\label{eqn:quantumconstraints}
\end{equation}
In the coordinate-representation we have
\begin{equation}
\bigg[-4 \kappa^2G_{ijkl} \frac{\delta^2}{\delta h_{ij} \delta
h_{kl}} + \frac{\sqrt{h}}{4 \kappa^2} \left( - \ {^3R}(h) + 2
\Lambda + 4 \kappa^2 \widehat{T}^{00} \right) \bigg]
\Psi[h_{ij},\Phi] =0 \,\, ,
\label{WdWeqnm}
\end{equation}
where $\langle h_{ij},\Phi | \Psi \rangle =  \Psi[h_{ij},\Phi]$ and $\widehat{T}^{00}=-\frac{1}{2 h} \frac{\delta^2}{
\delta \Phi^2} + \frac{1}{2} h^{ij} \Phi_{,i} \Phi_{,j} + V(\Phi)$. This constraint is called the Wheeler-DeWitt
equation. For pure gravity we have
\begin{equation}
\bigg[-4 \kappa^2G_{ijkl} \frac{\delta^2}{\delta h_{ij} \delta h_{kl}} + \frac{\sqrt{h}}{4 \kappa^2} \left( - \ {^3R}(h)
+ 2 \Lambda\right) \bigg] \Psi[h_{ij}] =0,
\end{equation}
where $\langle h_{ij}| \Psi \rangle =  \Psi[h_{ij}].$ In the general case the Wheeler-DeWitt equation is not amenable to extract physical information of the system. In order to obtain useful information it is a common practice to reduce the number of degrees of freedom. In the following sections we will consider models with one and two degrees of freedom.

\section{Deformation Quantization of Wheeler's Phase-Superspace}

In this section we work in a flat superspace with flat metric
$G_{ijkl}$. This case will allow us to introduce
the Weyl-Wigner-Groenewold-Moyal (WWGM) formalism for gravitational systems in
a direct way since the existence of the Fourier transform is warranted.
The more complicated cases of curved (mini)superspaces will be left
for a future work. The deformation quantization of gravity in ADM formalism and constrained systems is described in more detail in Refs. \cite{Antonsen:1997yq}.
We want to point out that in this section the calculations are formal just as is the case of path integrals in field theory and in order to obtain some physical results additional considerations need to be implemented depending on the specific system.

\subsection{The Stratonovich-Weyl Quantizer}

Let $F[h_{ij},\pi^{ij}; \Phi,\pi_{\Phi}]$ be a functional on the
phase space $\Gamma^*$ (Wheeler's phase superspace) and let
$\widetilde{F}[\mu^{ij},\lambda_{ij};\mu,\lambda]$ be its Fourier
transform. By analogy to the quantum mechanics case, we define the Weyl quantization rule as follows \cite{weyl,wigner,groene,moyal,vey,bayen,prueba,fedosov,k,Antonsen:1997yq,zachosrev,reviewone,reviewtwo,stern,sw,wwmformalism,tata,campos}
\small
\begin{equation}
\widehat{F}= {\cal W}(F[h_{ij},\pi^{ij}; \Phi,\pi_{\Phi}]) := \int {\cal D} \left(\frac{\lambda_{ij}}{2 \pi}\right) {\cal
D}\left(\frac{\mu^{ij}}{2 \pi}\right) {\cal D} \left(\frac{\lambda}{2 \pi}\right) {\cal D}\left(\frac{\mu}{2 \pi}\right) \widetilde{F}[\mu^{ij},\lambda_{ij};\mu,\lambda] \widehat{\cal U}[\mu^{ij},\lambda_{ij};\mu,\lambda], \label{fop}
\end{equation}
\normalsize
where the functional measures are given by ${\cal D}h_{ij} = \prod_{x} d h_{ij}(x)$ etc,  $\{ \widehat{\cal U}[\mu^{ij},\lambda_{ij};\mu,\lambda]: (\mu^{ij},\lambda_{ij};\mu,\lambda) \in \Gamma^* \}$ is the family of unitary operators given by
\begin{equation}
\widehat{\cal U}[\mu^{ij},\lambda_{ij};\mu,\lambda]:= \exp \bigg\{i\int dx \bigg( \mu^{ij}({x}) \widehat{h}_{ij}({x}) + \lambda_{ij}({x})
\widehat{\pi}^{ij}({x})  + \mu({x})
\widehat{\Phi}({x}) + \lambda({x}) \widehat{\pi}_{\Phi}({x})  \bigg)\bigg\},
\end{equation}
with $\widehat{h}_{ij}$, $\widehat{\pi}^{ij}$, $\widehat{\Phi}$ and $\widehat{\pi}_{\Phi}$ being the field operators defined as
$$
\widehat{h}_{ij}({x}) |h_{ij},\Phi \rangle = h_{ij}({x})
|h_{ij},\Phi \rangle \ \ \ \ {\rm and} \ \ \ \widehat{\pi}_{ij}({x})
|\pi_{ij},\pi_{\Phi} \rangle = \pi_{ij}({x}) |\pi_{ij},
\pi_{\Phi}\rangle ,
$$
\begin{equation}
\widehat{\Phi}({x}) |h_{ij},\Phi \rangle = \Phi(x)|h_{ij},\Phi \rangle \ \ \ \ {\rm and} \ \ \ \widehat{\pi}_{\Phi}({x})
| \pi_{ij}, \pi_{\Phi} \rangle = \pi_{\Phi}({x}) |\pi_{ij}, \pi_{\Phi} \rangle \,\,.
\end{equation}
These states form basis satisfying the completeness relations while operators satisfy the usual commutation rules. Using the well known Campbell-Baker-Hausdorff formula, the completeness relations and the standard commutation rules we can write $\widehat{\cal U}[\mu^{ij},\lambda_{ij};\mu,\lambda]$
in the following form
\begin{eqnarray}
\widehat{\cal U}[\mu^{ij},\lambda_{ij};\mu,\lambda]  &=&
\int {\cal D} {h}_{ij} {\cal D} {\Phi} \exp \bigg \{i \int dx \bigg( \mu^{ij}({x}) h_{ij}({x})  + \mu({x}) \Phi({x})\bigg) \bigg \}
\nonumber \\ &{}& \times \bigg| h_{ij} - \frac{\hbar \lambda_{ij}}{2}, \Phi - \frac{\hbar \lambda}{2} \bigg\rangle
\bigg\langle h_{ij} + \frac{\hbar \lambda_{ij}}{2}, \Phi + \frac{\hbar \lambda}{2}\bigg|. \label{unitary1}
\end{eqnarray}

It is easy to show, from (\ref{unitary1}), that one can obtain the following properties:
\begin{equation}
{\rm Tr} \bigg \{ \widehat{\cal U}
[\mu^{ij},\lambda_{ij};\mu,\lambda] \bigg \} = \delta [\lambda_{ij}]
\delta \left[\frac{\hbar \mu^{ij}}{2 \pi}\right] \cdot \delta
[\lambda] \delta \left[\frac{\hbar \mu}{2 \pi}\right],
\end{equation}
\begin{equation}
{\rm Tr} \bigg \{ \widehat{\cal U}^{\dagger} [\mu^{ij},\lambda_{ij};\mu,\lambda] \cdot \widehat{\cal U} [\mu'^{ij},\lambda'_{ij};\mu',\lambda'] \bigg \} = \delta [\lambda_{ij} - \lambda'_{ij}]\delta \left[ \frac{\hbar}{2 \pi}(\mu^{ij} -
\mu'^{ij})\right]  \cdot  \delta [\lambda - \lambda']\delta \left[ \frac{\hbar}{2 \pi}(\mu - \mu')\right], \label{trace2u}
\end{equation}
where ${\rm Tr}$ is the trace in some representation.

Equations (\ref{unitary1}) and (\ref{fop}) lead to write
\begin{equation}
\widehat{F} = \int {\cal D} \left(\frac{\pi^{ij}}{2 \pi
\hbar}\right) {\cal D} h_{ij}  {\cal D} \left(\frac{\pi_{\Phi}}{2
\pi \hbar}\right) {\cal D} \Phi F[h_{ij},\pi^{ij}; \Phi,\pi_{\Phi}]
\widehat{\Omega} [h_{ij},\pi^{ij}; \Phi,\pi_{\Phi}],
\end{equation}
where the operator $\widehat{\Omega}$ is given by
\small
\begin{align}
\widehat{\Omega} [h_{ij},\pi^{ij}; \Phi,\pi_{\Phi}] &= \int {\cal D} \left(\frac{\hbar \lambda_{ij}}{ 2 \pi}\right)  {\cal D}
\mu^{ij} {\cal D} \left(\frac{\hbar \lambda}{2 \pi}\right)  {\cal D} \mu \nonumber \\ \times \exp \bigg \{ -i \int dx
\bigg(\mu^{ij}(x) h_{ij} (x) + \lambda_{ij}(&x)\pi^{ij} (x) + \mu(x) \Phi(x) + \lambda(x) \pi_{\Phi}(x) \bigg) \bigg \}
\widehat{\cal U} [\mu^{ij},\lambda_{ij};\mu,\lambda].
\label{unitary2}
\end{align}
\normalsize
It is evident that the operator $\widehat{\Omega}$ is the Stratonovich-Weyl quantizer.  One can easily check the following properties of
$\widehat{\Omega}$
$$
\big (\widehat{\Omega} [h_{ij},\pi^{ij}; \Phi,\pi_{\Phi}] \big)^{^\dagger} = \big(\widehat{\Omega}[h_{ij},\pi^{ij}; \Phi,\pi_{\Phi}] \big),
$$
$$
{\rm Tr} \big \{ \widehat{\Omega}[h_{ij},\pi^{ij}; \Phi,\pi_{\Phi}] \big \} = 1,
$$
\begin{equation}
{\rm Tr} \bigg \{ \widehat{\Omega}[h_{ij},\pi^{ij}; \Phi,\pi_{\Phi}] \cdot
\widehat{\Omega}[h'_{ij},\pi'^{ij}; \Phi',\pi'_{\Phi}] \bigg \} = \delta \left[\frac{\pi^{ij} - \pi'^{ij}}{2 \pi \hbar}\right] \delta
[h_{ij} - h'_{ij}] \cdot \delta \left[\frac{\pi_{\Phi}- \pi'_{\Phi}}{2 \pi \hbar}\right] \delta [\Phi- \Phi']. \label{trace2om}
\end{equation}
Now it is possible to obtain the inverse map of ${\cal W}$ by multiplying (\ref{fop}) by $\widehat{\Omega}[h_{ij},\pi^{ij}; \Phi,\pi_{\Phi}]$, taking the trace of both sides and
using (\ref{trace2u})  we get
\begin{equation}
F[h_{ij},\pi^{ij}; \Phi,\pi_{\Phi}] = {\rm Tr} \bigg \{ \widehat{\Omega} [h_{ij},\pi^{ij}; \Phi,\pi_{\Phi}] \widehat{F}
\bigg\}. \label{FFop}
\end{equation}

One can also express $\widehat{\Omega}[h_{ij},\pi^{ij}; \Phi,\pi_{\Phi}]$ in a very useful form by inserting (\ref{unitary1}) into (\ref{unitary2}). Thus one gets
$$
\widehat{\Omega}[h_{ij},\pi^{ij}; \Phi,\pi_{\Phi}] = \int {\cal D} \xi_{ij}  \int {\cal D} \xi\exp \bigg \{ - \frac{i}{\hbar} \int dx \bigg( \xi_{ij}(x) \pi^{ij}(x) + \xi(x)
\pi_{\Phi}(x) \bigg) \bigg \}
$$
\begin{equation}
\times \bigg|h_{ij} - \frac{\xi_{ij}}{2}, \Phi - \frac{\xi}{2} \bigg\rangle \bigg\langle h_{ij} + \frac{\xi_{ij}}{2},
\Phi + \frac{\xi}{2} \bigg|. \label{Omegaop}
\end{equation}

\subsection{ The Star-Product}

Now we define the Moyal $\star-$product. Let $F=F[h_{ij},\pi^{ij}; \Phi,\pi_{\Phi}]$
and $G=G[h_{ij},\pi^{ij}; \Phi,\pi_{\Phi}]$ be some functionals on $\Gamma^*$ that correspond to the field operators
$\widehat{F}$ and $\widehat{G}$ respectively, i.e. $F[h_{ij},\pi^{ij}; \Phi,\pi_{\Phi}]={\cal W}^{-1}(\widehat{F})={\rm
Tr} \big( \widehat{\Omega} [h_{ij},\pi^{ij}; \Phi,\pi_{\Phi}] \widehat{F} \big)$ and
$G[\pi^{ij},h_{ij};\pi_{\Phi},\Phi]={\cal W}^{-1}(\widehat{G}) = {\rm Tr} \big(
\widehat{\Omega}[\pi^{ij},h_{ij};\pi_{\Phi},\Phi] \widehat{G} \big)$. The functional corresponding to the product  $\widehat{F} \widehat{G}$ will be denoted by $(F \star
G)[h_{ij},\pi^{ij}; \Phi,\pi_{\Phi}]$ and after some long but direct calculations we have that
\begin{equation}
(F\star G)[h_{ij},\pi^{ij}; \Phi,\pi_{\Phi}]:= {\cal W}^{-1}(\widehat{F} \widehat{G})= {\rm Tr} \bigg \{
\widehat{\Omega}[h_{ij},\pi^{ij}; \Phi,\pi_{\Phi}] \widehat{F} \widehat{G} \bigg \},
\end{equation}
gives rise to
\begin{equation}
\big(F \star  G\big)[h_{ij},\pi^{ij}; \Phi,\pi_{\Phi}] = F[h_{ij},\pi^{ij}; \Phi,\pi_{\Phi}] \exp\bigg\{\frac{i\hbar}{2}
\buildrel{\leftrightarrow}\over {\cal P}\bigg\} G[h_{ij},\pi^{ij}; \Phi,\pi_{\Phi}],
\label{MstarF}
\end{equation}
where the operator $\buildrel{\leftrightarrow}\over {\cal P}$ is defined as follows
\begin{equation}
\buildrel{\leftrightarrow}\over {\cal P} := \int dx \bigg(\frac{{\buildrel{\leftarrow}\over {\delta}}}{\delta h_{ij}(x)}
\frac{{\buildrel{\rightarrow}\over {\delta}}}{\delta \pi^{ij}(x)} - \frac{{\buildrel{\leftarrow}\over {\delta}}}{\delta
\pi^{ij}(x)} \frac{{\buildrel{\rightarrow}\over{\delta}}}{\delta h_{ij}(x)}\bigg) +  \int dx
\bigg(\frac{{\buildrel{\leftarrow}\over {\delta}}}{\delta \Phi(x)} \frac{{\buildrel{\rightarrow}\over {\delta}}}{\delta
\pi_{\Phi}(x)} - \frac{{\buildrel{\leftarrow}\over {\delta}}}{\delta \pi_{\Phi}(x)} \frac{{\buildrel{\rightarrow}\over
{\delta}}}{\delta \Phi(x)}\bigg).
\end{equation}
This is precisely the Moyal $\star-$product.

\subsection{The Wigner Functional}

In the deformation quantization formalism the Wigner function plays a very important role and in the same way finally we are going to define the Wigner functional. Let $\widehat{\rho}$ be the density operator of a quantum state.
The functional $\rho_{W}[h_{ij},\pi^{ij}; \Phi,\pi_{\Phi}]$ corresponding to $\widehat{\rho}$ reads
\begin{eqnarray}
&&\rho_{W}[h_{ij},\pi^{ij}; \Phi,\pi_{\Phi}] = {\rm Tr} \bigg \{ \widehat{\Omega}[h_{ij},\pi^{ij}; \Phi,\pi_{\Phi}]
\widehat{\rho}\bigg \} \nonumber \\
&&= \int {\cal D} \bigg({\xi_{ij} \over 2\pi \hbar}\bigg) {\cal D} \bigg({\xi \over 2\pi \hbar}\bigg)
 \exp \bigg\{ - \frac{i}{\hbar} \int dx \
\bigg(\xi_{ij}(x) \pi^{ij}(x) + \xi(x) \pi_{\Phi}(x)\bigg) \bigg\} \nonumber \\
&& \times \bigg\langle h_{ij} + \frac{\xi_{ij}}{2}, \Phi + \frac{\xi}{2}
\bigg| \widehat{\rho} \bigg| h_{ij} - \frac{\xi_{ij}}{2}, \Phi -\frac{\xi}{2}  \bigg\rangle.
\label{Wignerfunc}
\end{eqnarray}
For a pure state of the system $\widehat{\rho} = |\Psi\rangle \langle \Psi |$, the equation (\ref{Wignerfunc}) gives
$$
\rho_{_W}[h_{ij},\pi^{ij}; \Phi,\pi_{\Phi}] = \int {\cal D} \left(\frac{\xi_{ij}}{2 \pi \hbar}\right){\cal D} \left(\frac{\xi}{2 \pi
\hbar}\right) \exp \bigg\{ - \frac{i}{\hbar} \int dx \ \bigg( \xi_{ij}(x) \pi^{ij}(x) + \xi(x) \pi_{\Phi}(x) \bigg) \bigg\}
$$
\begin{equation}
\times \Psi^* \left[h_{ij} - \frac{\xi_{ij}}{2}, \Phi - \frac{\xi}{2} \right] \Psi \left[h_{ij} + \frac{\xi_{ij}}{2}, \Phi +
\frac{\xi}{2}\right],
\end{equation}
where $\langle h_{ij}, \Phi | \Psi \rangle = \Psi[h_{ij},\Phi]$ is the wave function of the universe.

The expected value of an arbitrary operator $\widehat{F}$ can then be obtained by means of $\widehat{\rho}$ as
\begin{eqnarray}
\langle\widehat{F}\rangle &=& \frac{{\rm Tr}(\widehat{\rho}\widehat{F})}{Tr(\widehat{\rho})} \nonumber \\
&=& \frac{\int {\cal D} \left(\frac{\pi^{ij}}{2\pi \hbar} \right){\cal D}h_{ij} {\cal D} \left(\frac{\pi_{\Phi}}{2\pi\hbar}\right){\cal D} \Phi \rho_{_W}[h_{ij},\pi^{ij}; \Phi,\pi_{\Phi}] {\rm Tr}(\widehat{\Omega}[h_{ij},\pi^{ij}; \Phi,\pi_{\Phi}] \widehat{F})}
{\int {\cal D} \pi^{ij} {\cal D}h_{ij} {\cal D} \pi_{\Phi} {\cal D} \Phi \rho_{_W}[h_{ij},\pi^{ij}; \Phi,\pi_{\Phi}]}.
\end{eqnarray}
It is now possible to write the equivalent of the constraints equations
(\ref{eqn:quantumconstraints}) in terms of the $\star -$product and the Wigner functional as
\begin{equation}
{\cal H}_{\perp} \star \rho_{_W}[h_{ij},\pi^{ij}; \Phi,\pi_{\Phi}] = 0,
\label{QHconstraint}
\end{equation}
\begin{equation}
{\cal H}_{i} \star \rho_{_W}[h_{ij},\pi^{ij}; \Phi,\pi_{\Phi}] = 0,
\label{QVconstraint}
\end{equation}
where ${\cal H}_{\perp}$ and ${\cal H}_{i}$ are given by (\ref{eqn:constraints1})
and (\ref{eqn:constraints2}) respectively. We term Eq. (\ref{QHconstraint}) the Wheeler-DeWitt-Moyal equation. It is important to mention that the dynamics is completely governed by these equations and that their explicit form depends of the particular system considered. These equations will be useful in the cosmological models which we will discuss in the following section. Also, we want to note that the real and imaginary parts of Eq. (\ref{QHconstraint}), encoded in the $\star -$product, are the generalized mass condition and the Wheeler-DeWitt-Vlasov transport equation presented in  \cite{Calzetta:1989vk} for flat minisuperspace.


\section{Deformation quantization in the minisuperspace}

Now we proceed to study the application of the deformation quantization construction to some models
in quantum cosmology. The general construction exposed in the previous section in terms of functional integrals is defined in the whole  superspace of 3-metrics and is necessary in a general case , however as a first approach we will deal with some models in the minisuperspace due to their simplicity and because they have been widely studied in the literature by other methods. Our intention is to obtain the Wigner function for these important cases and to motivate a further study using deformation quantization. We consider that this first step is necessary in order to gain some experience to eventually apply the deformation quantization formalism to curved phase-space. It is known that this formalism has a natural extension to these situations (see \cite{fedosov,k}) and allows another suitable extensions or generalizations.

\subsection{de Sitter cosmological model}

As a first example we will deal with a minisuperspace model where the phase space is
bidimensional. We are going to calculate the Wigner function from the Wheeler-DeWitt-Moyal equation
and also from its integral expression. Let start with a minisuperspace (where the degrees of freedom are reduced to just one represented by the scale factor of the universe) corresponding to the Friedmann-Robertson-Walker metric
\begin{equation}
ds^{2}= l_{p}^{2}\left[-N^2 dt^{2}+ a^{2}(t)d\Omega^{2}_{3}\right],
\end{equation}
where $a$ is the scale factor of the universe, $N$ is the lapse function, $d\Omega^{2}_{3}$ is the metric of the unit three-sphere, $l_{p}=2/3 L_{p}$ and $L_p$ denotes the Planck
length. Introducing new variables $q=a^2$ and $\tilde{N}= qN$  \cite{Luoko} the Hamiltonian ${\cal H}_{\perp}$ cast out in an easy form
\begin{equation}
{\cal H}_{\perp} = \frac{1}{2}\left( -4p^2 + \lambda q -1 \right),
\end{equation}
where $\lambda$ is the cosmological constant in Planck units.  In the coordinate representation the Hamiltonian acquires a simple form and its dynamics
corresponds to a particle in a linear potential. The Wheeler-DeWitt equation comes from the Hamiltonian constraint (\ref{WdWeqnm}) and is written as
\begin{equation}
\left( 4\hslash ^{2}\frac{d^{2}}{dq^{2}} + \lambda q - 1 \right)\Psi(q)= 0\,.
\end{equation}
Depending of the boundary conditions it can be found \cite{approch} the Vilenkin wave function
\begin{equation}
\Psi_{V} (q) = \frac{1}{2}\mbox{Ai}\left( \frac{1-\lambda q}{(2\lambda \hslash )^{2/3}} \right) + \frac{i}{2}\mbox{Bi}\left(
\frac{1-\lambda q}{(2\lambda    \hslash )^{2/3}} \right),
\end{equation}
the Hartle-Hawking wave function
\begin{equation}
\Psi_{HH}(q)= \mbox{Ai}\left( \frac{1-\lambda q}{(2\lambda \hslash )^{2/3}} \right)\, ,
\end{equation}
and the Linde wave function
\begin{equation}
\Psi_{L}(q)= -i\mbox{Bi}\left( \frac{1-\lambda q}{(2\lambda \hslash )^{2/3}} \right)\, ,
\end{equation}
where $\mbox{Ai}(x)$ and $\mbox{Bi}(x)$ are the Airy functions of first and second class respectively.

One of the main differences between the Hartle-Hawking, Linde and Vilenkin wave functions is their behavior in the region $q>1/\lambda$. The Vilenkin tunneling wave function has the following expression
\begin{equation}
 \Psi_{V} = \psi_{-}(q),
\end{equation}
the Hartle-Hawking wave function is
\begin{equation}
\Psi_{HH} = \psi_{+}(q) + \psi_{-}(q),
\end{equation}
and the Linde wave function is written as
\begin{equation}
\Psi_{L} = \psi_{+}(q) - \psi_{-}(q),
\end{equation}
where $\psi_{-}(q)$, $\psi_{+}(q)$ describe an expanding and contracting universe, respectively.
In the classical allowed region $q>1/\lambda$  the WKB solutions are
\begin{equation}
\psi_{\pm} = [p(q)]^{-1/2} \exp\left[\pm i \int^q _{1/\lambda} p(q')dq' \mp \frac{i\pi}{4}\right],
\label{eqn:wkb}
\end{equation}
where $p(a) = [-U(a)]^{1/2}=\frac{1}{2}[\lambda q-1]^{1/2}$. From the WKB wave function (\ref{eqn:wkb}) we have
\begin{equation}
{\hat p} \psi_{\pm}(q) \approx  \pm p \psi_{\pm}(q).
\label{eqn:momentumwkb}
\end{equation}
Taking into account that
\begin{equation}
p = - \frac{{\dot q}}{4 \tilde{N}},
\end{equation}
the Eq. (\ref{eqn:momentumwkb}) confirms the already given interpretation for $\psi_{\pm}(q)$, i.e. negative values of $p$ correspond to an expanding universe. Therefore, the Vilenkin wave function includes only an expanding component while the Hartle-Hawking and Linde wave functions include expanding and contracting universes with equal weight (for a different interpretation see \cite{Rubakovint}).
In fact, for large values of $q$ these wave functions have the following expressions
\begin{equation}
 \Psi_{V}(q) \approx \frac{(2\lambda \hbar)^{1/6}}{2[\pi^{2}(\lambda q -1)]^{1/4}}e^{-iS},
\end{equation}
\begin{equation}
 \Psi_{HH}(q) \approx \frac{(2\lambda \hbar)^{1/6}}{[\pi^{2}(\lambda q -1)]^{1/4}}\cos S = \frac{(2\lambda \hbar)^{1/6}}{2[\pi^{2}(\lambda q -1)]^{1/4}}\left(e^{iS} + e^{-iS}\right),
\end{equation}
\begin{equation}
 \Psi_{L}(q) \approx i\frac{(2\lambda \hbar)^{1/6}}{[\pi^{2}(\lambda q -1)]^{1/4}}\sin S = \frac{(2\lambda \hbar)^{1/6}}{2[\pi^{2}(\lambda q -1)]^{1/4}}\left(e^{iS} - e^{-iS}\right),
\end{equation}
where
\begin{equation}
 S= \frac{1}{3\lambda \hbar}(\lambda q -1)^{3/2}-\frac{\pi}{4}.
\end{equation}
Now we proceed to calculate the Wigner function directly by solving the Wheeler-DeWitt-Moyal equation (\ref{QHconstraint})
\begin{equation}
 {\cal H}_{\perp} \star \rho_{W} = 0.
\label{hamiltonian}
\end{equation}
where the corresponding Moyal $\star-$product is given by
\begin{equation}
 \star = \exp \left\lbrace \frac{i\hslash}{2} \buildrel{\leftrightarrow}\over {\cal P}\right\rbrace = \exp \left\lbrace \frac{i\hslash}{2} \left( \frac{{\buildrel{\leftarrow}\over {\partial}}}{\partial q} \frac{{\buildrel{\rightarrow}\over {\partial}}}{\partial
p} - \frac{{\buildrel{\leftarrow}\over {\partial}}}{\partial p} \frac{{\buildrel{\rightarrow}\over
{\partial}}}{\partial q}\right) \right\rbrace .
\end{equation}
Taking in consideration the exponential power series expansion we have
\begin{equation}
\frac{1}{2}(- 4p^2 + \lambda q -1)\left[ \sum_{n=0} ^{\infty} \frac{1}{n!}\left( \frac{i\hslash}{2}\right) ^{n}
\buildrel{\leftrightarrow}\over {\cal P} ^{n} \right]\rho_{W}=0 \,\, ,
\end{equation}
which can be written as
\begin{equation}
(-4p^2 + \lambda q -1) \rho_{W} + \frac{i\hslash}{2}\lambda\partial_{p}\rho_{W} +4i\hslash p \partial_{q}\rho_{W} +
\hslash^{2}\partial^{2}_{q}\rho_{W}= 0. \label{lastdefeq}
\end{equation}
If we define a new variable $ z= 4p^2 - \lambda q + 1$ the imaginary part of the former equation is identically zero and for the real part we obtain a new form of the equation (\ref{lastdefeq})
\begin{equation}
\hslash^{2}\lambda^{2} \frac{d^2 \rho_{W}}{dz^2} - z\rho_{W}= 0 \,\, ,
\end{equation}
whose solution is
\begin{equation}
\rho_W = c_1\mbox{Ai}\left( \frac{4p^2-\lambda
q+1}{(\hslash\lambda)^{2/3}}\right). \label{rhowigner}
\end{equation}
In order to obtain the last result the formalism assumes the existence of the Wigner transform, where the range of integration in this transform is from minus to plus infinity in the variable $q$. However, the valid range interval for the variable $q$ is positive but still the Wigner function is very similar to (\ref{rhowigner}) because of the exponential decay for positive values of the argument of Ai$(x)$. Last equation indeed admits another solution corresponding to $\mbox{Bi}\left( \frac{4p^2-\lambda q+1}{(\hslash\lambda)^{2/3}}\right)$, but in this case, the Airy function $\mbox{Bi}(x)$ is divergent for positive $x$ and this part cannot be included as a suitable Wigner function. In fact, in order to obtain an appropriate solution from (\ref{hamiltonian}), the potential could be complemented by an infinite barrier avoiding the existence of negatives values for $q$. However the implementation of this procedure is very cumbersome as it has been shown in \cite{Dias1} and \cite{Dias2} and it is not convenient to develop it here.

\begin{figure}
\begin{minipage}[t]{8cm}
\begin{center}
\includegraphics[scale=0.68]{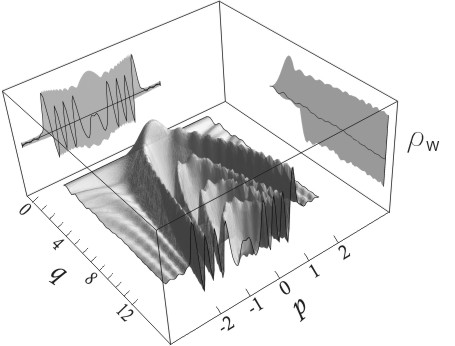}
\caption[Short caption for figure 1]{\label{fig:WHH} {\scriptsize The Hartle-Hawking Wigner function ($\hbar= \lambda=1$). The figure shows many oscillations due to the interference between wave functions of expanding and contracting universes.}}
\end{center}
\end{minipage}
\hfill
\begin{minipage}[t]{8cm}
\begin{center}
\includegraphics[scale=0.48]{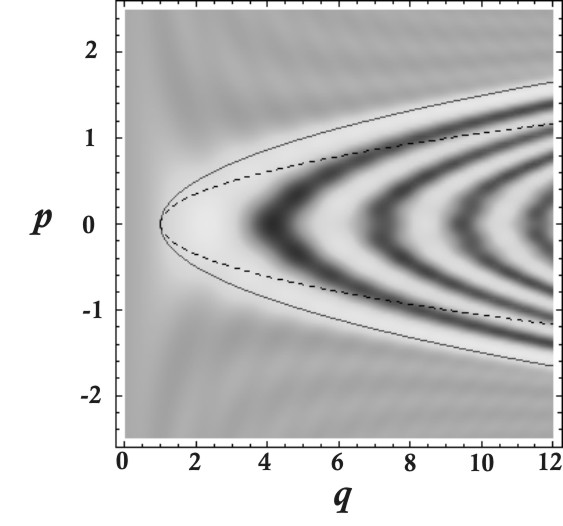}
\caption[Short caption for figure 2]{\label{fig:WHH2} {\scriptsize Hartle-Hawking Wigner function density projection. It can be observed that the classical trajectory does not coincide with the highest peaks of the Wigner function.}}
\end{center}
\end{minipage}
\end{figure}

Instead of this we can calculate the Wigner function using the following integral expression \cite{wigner}
\begin{equation}
\rho_{W}(q, p) = \int^{\infty} _{-\infty} \frac{d\xi}{2\pi \hslash} \exp \bigg\{-i \frac{\xi}{\hslash}p \bigg\} \Psi^*\left( q- \frac{\xi}{2}\right) \Psi \left( q + \frac{\xi}{2}\right).
\label{intwigner}
\end{equation}
Employing the convolution theorem and the Fourier transform of the Airy function we get the Wigner function for the Hartle-Hawking wave function
\begin{equation}
\rho_W(q,p) = \frac{2^{1/3}}{\pi(\hbar \lambda)^{1/3}} \mbox{Ai}\left[\frac{4p^{2}-\lambda q + 1}{(\hbar \lambda)^{2/3}}\right].
\end{equation}
The last result was obtained integrating out from minus to plus infinity in the variable $q$ (in fact this result was already derived in \cite{Habib:1990hx}). However, the wave function is only valid defined for positive values of the scale factor and the $q$ variable. Due to the fact that the Airy function Ai$(x)$ presents an exponential decay for positive values of $x$, the Wigner function is very similar to the expression obtained in terms of the Airy function. If we restrict the range of integration of $q$ for positive values the computation for an analytical expression of the Wigner function is very complicated. This problem is similar as the one we faced for the Wheeler-DeWitt-Moyal equation.
We cope with this complication by implementing a Fortran code in order to calculate numerically the Wigner function. The result is given in Fig. \ref{fig:WHH} and Fig. \ref{fig:WHH2}, where the continuous line corresponds to the classical trajectory (\ref{traject1}) and the dashed line to the trajectory (\ref{traject2}).

\begin{figure}
\begin{minipage}[t]{8cm}
\begin{center}
\includegraphics[scale=0.55]{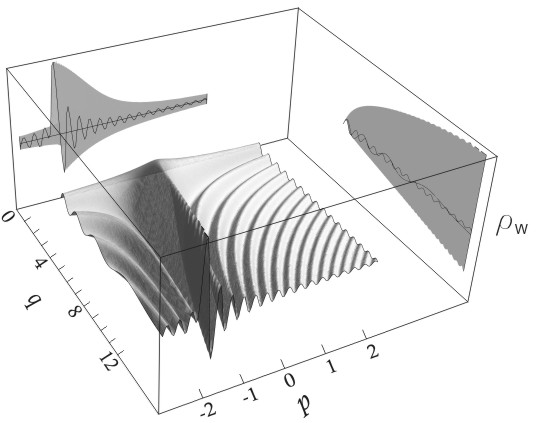}
\caption[Short caption for figure 1]{\label{fig:WV} {\scriptsize The Vilenkin Wigner function. It is observed a clear maximum and less oscillations compared with the Hartle-Hawking case.}}
\end{center}
\end{minipage}
\hfill
\begin{minipage}[t]{8cm}
\begin{center}
\includegraphics[scale=0.48]{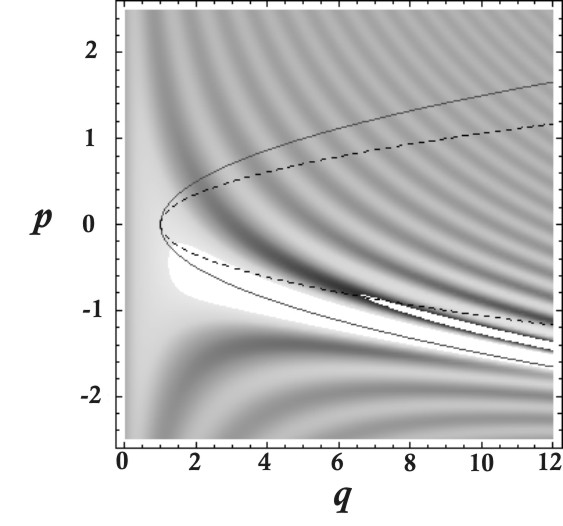}
\caption[Short caption for figure 2]{\label{fig:WV2} {\scriptsize The density projection of the Vilenkin  Wigner function. The classical trajectory is at some parts on the maxima of the Wigner function and has only one branch.}}
\end{center}
\end{minipage}
\end{figure}

\begin{figure}
\begin{minipage}[t]{8cm}
\begin{center}
\includegraphics[scale=0.68]{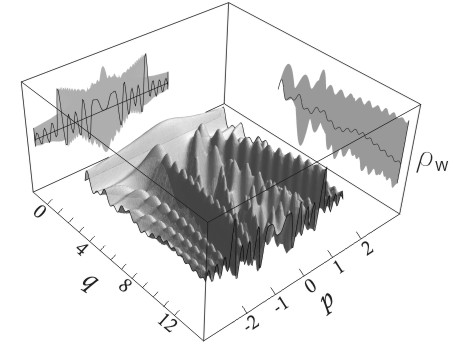}
\caption[Short caption for figure 1]{\label{fig:WL} {\scriptsize Wigner function for the Linde wave function. The figure shows a reduction in the amplitude of the oscillations compared to the Hartle-Hawking.}}
\end{center}
\end{minipage}
\hfill
\begin{minipage}[t]{8cm}
\begin{center}
\includegraphics[scale=0.5]{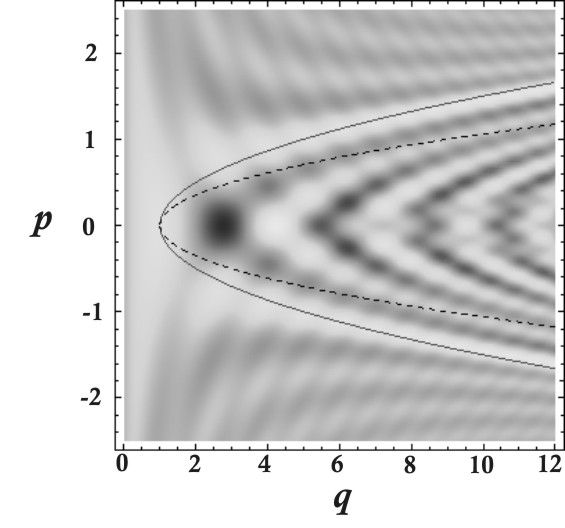}
\caption[Short caption for figure 2]{\label{fig:WL2} {\scriptsize The density projection of the Linde Wigner function. In this case the classical trajectory coincides with the highest peak of its corresponding Wigner function.}}
\end{center}
\end{minipage}
\end{figure}
For the Vilenkin wave function it is very difficult to get an analytical expression for the Wigner function directly from  (\ref{intwigner}) even using the WKB approximation for the wave function taking into account that the integration should be restricted for positive values of $q$. Again, we performed a numerical analysis to obtain the results that are depicted in Fig. \ref{fig:WV} and Fig. \ref{fig:WV2}, where the meaning of the continuous and dashed lines are the same as in the Hartle-Hawking case.

As before, the Linde Wigner function is very complicated to work with. We calculated the Wigner function numerically and its behavior is presented in Fig. \ref{fig:WL} and Fig. \ref{fig:WL2}.

These plots can be described using the WKB approximation for the wave function and the general statements presented in \cite{Habib:1990hx}. In terms of the WKB density matrix $\rho_{WKB}$ the evaluation of
\begin{equation}
\rho_{W}(q,p) \approx \frac{1}{\pi \hbar} \int^{\infty} _{-\infty} d\alpha \exp{(-2ip\alpha)}\rho_{WKB}(q+\alpha,q-\alpha)
\end{equation}
for the Hartle-Hawking wave function includes four terms of the form: $\exp(iS(q+\alpha) - iS(q-\alpha))$, $\exp(-iS(q+\alpha) - iS(q-\alpha))$, $\exp(iS(q+\alpha) + iS(q-\alpha))$, $\exp(-iS(q+\alpha) + iS(q-\alpha))$. These terms come from the products $\psi_{+}(q+\alpha) \psi_{+} ^{*} (q- \alpha)$, $\psi_{-}(q+\alpha) \psi_{+} ^{*} (q- \alpha)$, $\psi_{+}(q+\alpha) \psi_{-} ^{*} (q- \alpha)$, $\psi_{-}(q+\alpha) \psi_{-} ^{*} (q- \alpha)$, respectively. By means of the stationary phase approximation, each one of the former terms will contribute to the Wigner function in different regions bounded by the classical trajectory
\begin{equation}
 p^2 = \frac{1}{4}(\lambda q -1),
\label{traject1}
\end{equation}
and by the trajectory
\begin{equation}
 p^2 = \frac{1}{8}(\lambda q -1),
 \label{traject2}
\end{equation}
which corresponds to points where the path integral has a saddle point at zero momentum \cite{Habib:1990hx}, i.e. where the classical action changes sign. In the upper region of the phase space between the classical trajectory and the dashed curve the Wigner function has contributions mainly coming from the first term in the density matrix. The region inside the dashed curve receives the principal contributions form the second and third terms of the density matrix. The fourth term of the density matrix has a saddle point in the bottom part of phase space between the classical trajectory and the dashed curve and as a consequence it contributes predominantly in this region (see Fig. \ref{fig:WHH2}).

The Linde Wigner function has a similar structure like the Hartle-Hawking case but there is one important difference: the interference terms between contracting and expanding universes have an opposite sign with respect to Hartle-Hawking. The consequence of this difference is the reduction of the amplitudes of the oscillations inside the classical trajectory (see Fig. \ref{fig:WL}).

The Wigner function for the Vilenkin wave function only receives a contribution from density matrix corresponding to $\psi_{-}(q+\alpha) \psi_{-} ^{*} (q- \alpha)$ and it has a predominantly contribution only in the lower region of phase space between the classical trajectory and the dashed line. This behavior is reflected in Fig. \ref{fig:WV2}. It is important to mention that decoherence of the Vilenkin Wigner function appears to be easier taking into account that large amplitude terms due to interference between collapsing and expanding universes are not present near $p=0$.

We discuss now the interpretation of these numerical results.
For the Hartle-Hawking Wigner function we see an oscillatory behavior where the largest peaks are not on the classical trajectory but away from it by a distance of $O(\hbar^{(2/3)})$, we can appreciate this fact in the density plot. It should be remarked the existence of oscillations of the Wigner function that are not near of the classical trajectory, however the heights of these peaks decreases with the distance to the classical trajectory.

The Wigner function for the Linde wave function present a similar pattern, in general terms, like the Hartle-Hawking wave function, however it is possible to see some differences. We can appreciate more fluctuations of the Wigner function inside the region corresponding to the classical trajectory than the Hartle-Hawking Wigner function. Furthermore, the amplitude of the oscillations are smaller for the Linde wave function than for Hartle-Hawking. The largest peaks of the Wigner function correspond to the classical trajectory.

In the case of the Wigner function for the Vilenkin wave function the classical trajectory is at some parts of the curve at the middle of the largest peaks of the Wigner function, i.e. the Vilenkin Wigner function has the largest peaks more closely to the classical trajectory than the Hartle-Hawking Wigner function, but only in one part. This behavior is expected, as we explained above,  due to the fact that the Vilenkin wave function represents the tunneling wave function and as a consequence there is only one part of the classical trajectory corresponding to negative values of the momenta (expanding universe).


\subsection{Kantowski-Sachs model}


Another interesting case to deal with under the deformation quantization procedure is the cosmological Kantowski-Sachs model (KS) \cite{Kantowski}. We consider the metric in the Misner parametrization \cite{Misner}
\begin{equation}
ds^2 = -N^2dt^2 + e^{2\sqrt{3}\beta} dr^2 +  e^{-2\sqrt{3}\beta} e^{-2\sqrt{3}\Omega}( d\theta ^2 + \sin^2 \theta d\varphi^2).
\end{equation}
Choosing a particular factor ordering the Wheeler-DeWitt equation takes the following form
\begin{equation}
 \left(- P^2 _\Omega + P^2 _\beta - 48\exp(-2\sqrt{3}\Omega) \right) \Psi (\Omega,\beta) = 0,
\end{equation}
where $P_\Omega = -i\hbar\frac{\partial}{\partial \Omega}$ and $P_\beta = -i\hbar\frac{\partial}{\partial \beta}$. The solutions of the former equation are given by \cite{Misner}
\begin{equation}
\Psi_\nu ^{\pm} (\Omega,\beta)=e^{\pm i\nu \sqrt{3} \beta}K_{i\frac{\nu}{\hbar}} \left(\frac{4}{\hbar}e^{-\sqrt{3}\Omega}\right),
\end{equation}
where $K_{i\nu}(x)$ is the MacDonald function of imaginary order. The normalized wave function is
\begin{equation}
 \Psi_\nu (\Omega,\beta)= \frac{3^{1/4}}{\pi \hbar}\sqrt{\sinh\left(\frac{\pi \nu}{\hbar}\right)}K_{i\frac{\nu}{\hbar}} \left(\frac{4}{\hbar}e^{-\sqrt{3}\Omega}\right)
\end{equation}
which satisfy \cite{radoslaw}
\begin{equation}
\int_{-\infty} ^{\infty} d\Omega d \beta \Psi_{\nu} ^*(\Omega, \beta)\Psi_{\nu '}(\Omega, \beta)= \delta(\nu^2 -\nu '^2).
\end{equation}
In order to write the Wheeler-DeWitt-Moyal equation (\ref{QHconstraint}) it is very useful to employ the next relation
\begin{equation}
f(x,p) \star g(x,p) = f\left( x + \frac{i\hbar}{2}{\buildrel{\rightarrow}\over{\partial}}_p, p - \frac{i\hbar}{2}{\buildrel{\rightarrow}\over{\partial}}_x\right)g(x,p).
\label{shift}
\end{equation}
In this way, we obtain the following equation
\begin{equation}
\left(-\left( P_{\Omega} - \frac{i\hbar}{2}{\buildrel{\rightarrow}\over{\partial}}_\Omega\right)^2  + \left( P_{\beta} - \frac{i\hbar}{2}{\buildrel{\rightarrow}\over{\partial}}_\beta\right)^2 - 48e^{\left(-2\sqrt{3}\left(\Omega + \frac{i\hbar}{2}{\buildrel{\rightarrow}\over{\partial}}_{P_{\Omega}}\right)\right)}\right) \rho (\Omega,P_{\Omega},\beta, P_{\beta}) = 0,
\end{equation}
which can be split into two equations corresponding to its real part
\begin{equation}
\left[ -P_{\Omega}^{2} + \frac{\hbar^{2}}{4}\partial^{2}_{\Omega} + P_{\beta}^{2} - \frac{\hbar^{2}}{4}\partial^{2}_{\beta} - 48e^{-2\sqrt{3}\Omega}\cos \left(\sqrt{3}\hbar\partial_{P_{\Omega}}\right)\right]\rho = 0,
\label{real}
\end{equation}
and its imaginary part
\begin{equation}
\left[ \hbar(P_\Omega \partial_\Omega) - \hbar(P_\beta \partial_\beta) + 48 e^{-2\sqrt{3}\Omega} \sin\left(\sqrt{3}\hbar \partial_{P_\Omega}\right)\right] \rho = 0.
\label{imaginary}
\end{equation}
If we propose $\rho(\Omega, P_{\Omega}, \beta, P_{\beta}) = \rho_\Omega(\Omega, P_{\Omega}) \rho_\beta(\beta, P_{\beta})$, and taking into account that $e^{i\sqrt{3}\hbar \partial_x}f(x) = f(x+ i\sqrt{3}\hbar)$ and also that for free particle in the $\beta$ parameter $\partial_\beta \rho_\beta = 0$, we get from equation (\ref{imaginary})
\begin{eqnarray}
\nonumber
\partial^2 _\Omega \rho_\Omega = -\frac{48\sqrt{3}i}{\hbar P_\Omega}e^{-2\sqrt{3}\Omega}( \rho(\Omega, P_\Omega + i\sqrt{3} \hbar) - \rho(\Omega, P_\Omega - i\sqrt{3} \hbar)) \\ \nonumber
- \frac{576}{\hbar^2}\frac{e^{-2\sqrt{3}\Omega}}{P_\Omega ^2 + 3\hbar ^2}\bigg[ \rho(\Omega, P_\Omega + 2i\sqrt{3}\hbar) - 2\rho(\Omega, P_\Omega)
+  \rho(\Omega, P_\Omega - 2i\sqrt{3}\hbar)  \\
- \frac{i\sqrt{3}\hbar}{P_\Omega}(\rho(\Omega, P_\Omega + 2i\sqrt{3}\hbar) - \rho(\Omega, P_\Omega - 2i\sqrt{3}\hbar))\bigg] .
\end{eqnarray}
Using the last equation in (\ref{real}) we finally obtain
\begin{eqnarray}
\nonumber
&-&P_\Omega ^2 \rho_\Omega - \frac{12\sqrt{3}i\hbar^2 e^{-2\sqrt{3}\Omega}}{P_\Omega}\left[\rho_\Omega(\Omega, P_\Omega + i\sqrt{3}\hbar) - \rho_\Omega(\Omega, P_\Omega - i\sqrt{3}\hbar)\right]\\ \nonumber
&-&\frac{144\hbar^2 e^{-4\sqrt{3}\Omega}}{(P_\Omega ^2 +3\hbar^2)}\left[\rho_\Omega(\Omega, P_\Omega + 2i\sqrt{3}\hbar) - 2\rho_\Omega(\Omega, P_\Omega)+ \rho_\Omega(\Omega, P_\Omega - 2i\sqrt{3}\hbar)\right]  \\  \nonumber
&+&\frac{144\sqrt{3}i\hbar^3 e^{-4\sqrt{3}\Omega}}{P_\Omega (P_\Omega ^2 +3\hbar^2)}\left[\rho_\Omega(\Omega, P_\Omega + 2i\sqrt{3}\hbar) - \rho_\Omega(\Omega, P_\Omega - 2i\sqrt{3}\hbar)\right]\\
&-& 24 e^{-2\sqrt{3}\Omega} \left[\rho_\Omega(\Omega, P_\Omega + i\sqrt{3}\hbar) +  \rho_\Omega(\Omega, P_\Omega - i\sqrt{3}\hbar)\right] = -P_\beta ^2 \rho_\Omega.
\label{KSfinal}
\end{eqnarray}
It is hard to obtain directly a solution to this equation, so we will follow a different approach and will use the integral representation for the Wigner function.
Using the following result (see Sec. 19.6 formula (25) in \cite{ErdelyiTables} and the comment in \cite{CFZ})
\begin{equation}
\int_0 ^{\infty} dw(wz)^{1/2} w^{\sigma -1} K_\mu (a/w) K_\nu (wz) = 2^{-\sigma -5/2} a^\sigma G^{40}_{04}\left(\frac{a^2 z^2}{16} {\bigg |} \frac{\mu - \sigma}{2}, \frac{-\mu - \sigma}{2}, \frac{1}{4}+ \frac{\nu}{2}, \frac{1}{4}- \frac{\nu}{2}\right),
\end{equation}
where $G^{40}_{04}\left(\frac{a^2 z^2}{16} {\bigg |} \frac{\mu - \sigma}{2}, \frac{-\mu - \sigma}{2}, \frac{1}{4}+ \frac{\nu}{2}, \frac{1}{4}- \frac{\nu}{2}\right)$ is a special case of Meijer's $G$ function (see Sec. 5.3 in \cite{Erdelyi})
\begin{equation}
 G^{mn} _{pq} \left( z \bigg | \begin{array}{lc}
a_{i}, & i=1,...,p \\
b_{j}, & j=1,...,q
\end{array}
\right),
\end{equation}
we calculate the Wigner function for the $\Omega$ part
\begin{equation}
\rho_\Omega (\Omega, P_\Omega) = \frac{3^{1/2}}{2\pi^4 \hbar^2 }\sinh\left(\frac{\pi \nu}{\hbar}\right)\int_{-\infty} ^{\infty} dy K_{i\frac{\nu}{\hbar}} \left(\frac{4}{\hbar}e^{-\sqrt{3}\left(\Omega -\frac{\hbar}{2}y\right)}\right) e^{-iyP_\Omega} K_{i\frac{\nu}{\hbar}} \left(\frac{4}{\hbar}e^{-\sqrt{3}\left(\Omega +\frac{\hbar}{2}y\right)}\right).
\label{roks}
\end{equation}
Then, we obtain the following expression for the Wigner function
\begin{align}
\nonumber
 \rho_\Omega (\Omega, P_\Omega) & =  \frac{\sinh(\pi \nu/\hbar)}{\pi^3} \frac{e^{\sqrt{3}\Omega}}{16\hbar^{2}}
 \left(\frac{2}{\hbar^{2}}e^{-\sqrt{3}\Omega}\right)^{-\frac{2iP_{\Omega}}{\sqrt{3}\hbar}} \\
 & \times G^{40}_{04}
 \left( \frac{16}{\hbar^{4}}e^{-4\sqrt{3}\Omega}\bigg | \frac{1}{4} +i \left(\frac{\nu}{2\hbar} + \frac{P_\Omega}{\sqrt{3}\hbar}\right),\frac{1}{4} +i \left(\frac{-\nu}{2\hbar} + \frac{P_\Omega}{\sqrt{3}\hbar}\right), \frac{1}{4} + \frac{i\nu}{2\hbar}, \frac{1}{4} - \frac{i\nu}{2\hbar} \right).
\label{eqn:WignerKS}
\end{align}
Now, employing the Meijer's function property
\begin{equation}
 x^{\sigma}G^{mn}_{pq}\left( x \bigg | \begin{array}{lc}
a_{i}, & i=1,...,p \\
b_{j}, & j=1,...,q
\end{array}
\right) = G^{mn}_{pq}\left( x \bigg | \begin{array}{lc}
a_{i} +\sigma, & i=1,...,p \\
b_{j}+ \sigma, & j=1,...,q
\end{array}
\right),
\end{equation}
the equation (\ref{eqn:WignerKS}) can be written in the following form
\footnotesize
\begin{align}
\nonumber
 &\rho_\Omega (\Omega, P_\Omega) = \frac{\sinh(\pi \nu/\hbar)}{\pi^3}\frac{e^{\sqrt{3}\Omega}}{16\hbar^{2}} \\
 & \times G^{40}_{04}\left(\frac{16}{\hbar^{4}}e^{-4\sqrt{3}\Omega}\bigg | \frac{1}{4} +i \left(\frac{\nu}{2\hbar} + \frac{P_\Omega}{2\sqrt{3}\hbar}\right),\frac{1}{4} +i \left(\frac{-\nu}{2\hbar} + \frac{P_\Omega}{2\sqrt{3}\hbar}\right), \frac{1}{4} + i \left( \frac{\nu}{2\hbar} - \frac{P_\Omega}{2\sqrt{3}\hbar}\right), \frac{1}{4} +i \left(\frac{-\nu}{2\hbar} - \frac{P_\Omega}{2\sqrt{3}\hbar}\right) \right).
\end{align}
\normalsize
It is possible to verify that the Wigner function indeed satisfy equation (\ref{KSfinal}).

\begin{figure}
\begin{minipage}[t]{8cm}
\begin{center}
\includegraphics[scale=0.65]{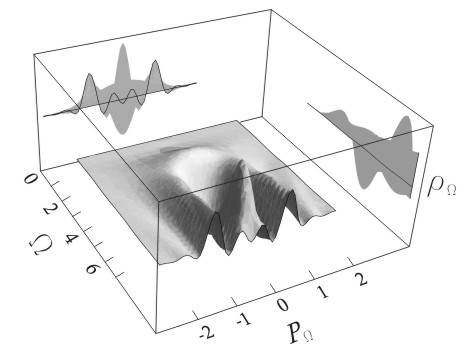}
\caption[Short caption for figure 1]{\label{fig:WKS1} {\scriptsize The Kantowski-Sachs Wigner function for $\nu = 0.5$. Few oscillations are present for this case with a clear maximum around $\Omega = 6$ and $P_{\Omega}=0$.}}
\end{center}
\end{minipage}
\hfill
\begin{minipage}[t]{8cm}
\begin{center}
\includegraphics[scale=0.47]{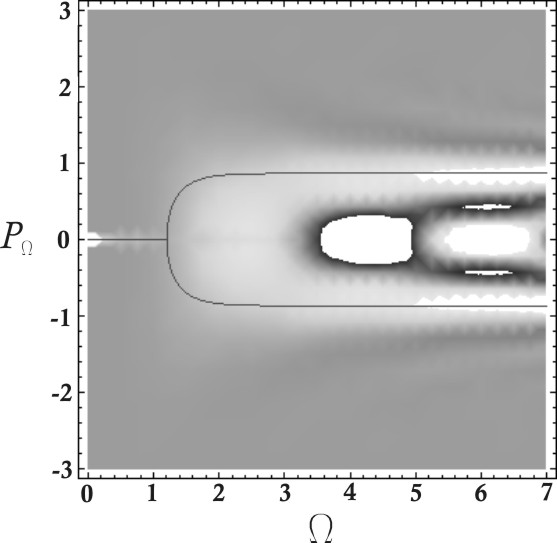}
\caption[Short caption for figure 2]{\label{fig:WKS1Density} {\scriptsize Kantowski-Sachs Wigner function density projection for $\nu =0.5$. It can be observed that the value of the Wigner function is large in an ample area.}}
\end{center}
\end{minipage}
\end{figure}

\begin{figure}
\begin{minipage}[t]{8cm}
\begin{center}
\includegraphics[scale=0.65]{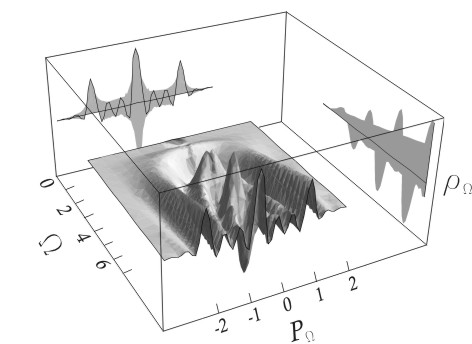}
\caption[Short caption for figure 1]{\label{FIGWKS2} {\scriptsize The Wigner function for the Kantowski-Sachs wave function for $\nu=1$. The number of oscillations are maxima around $P_{\Omega}=0$.}}
\end{center}
\end{minipage}
\hfill
\begin{minipage}[t]{8cm}
\begin{center}
\includegraphics[scale=0.47]{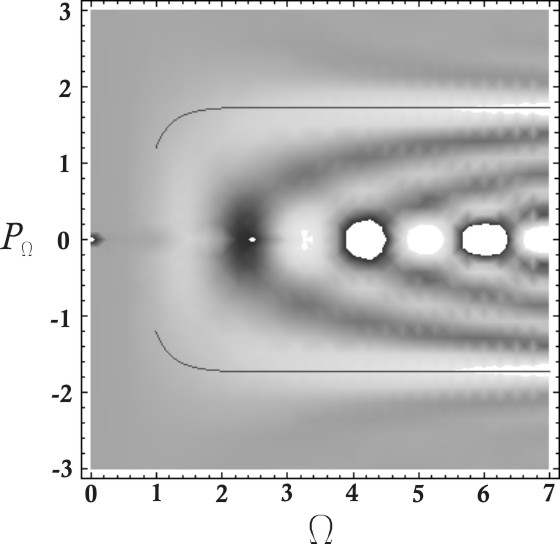}
\caption[Short caption for figure 2]{\label{fig:WKS2Density} {\scriptsize Density projection of the Kantowski-Sachs Wigner function for $\nu = 1$. It can be observed that classical trajectory is  close to the exterior peaks of the oscillations.}}
\end{center}
\end{minipage}
\end{figure}

\begin{figure}
\begin{minipage}[t]{8cm}
\begin{center}
\includegraphics[scale=0.65]{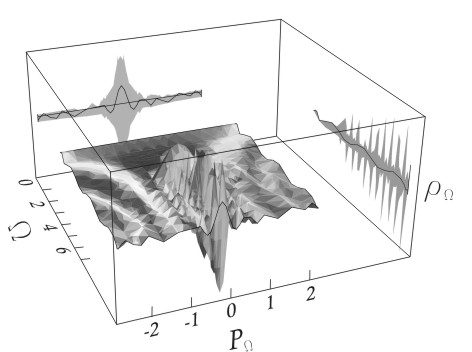}
\caption[Short caption for figure 1]{\label{fig:WKS3} {\scriptsize Kantowski-Sachs Wigner function for $\nu=4$. There is a considerable increase in the number of oscillations centered at $P_{\Omega}=0$.}}
\end{center}
\end{minipage}
\hfill
\begin{minipage}[t]{8cm}
\begin{center}
\includegraphics[scale=0.47]{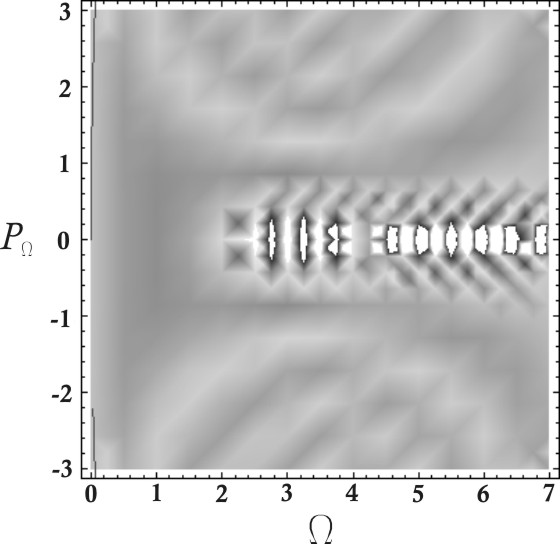}
\caption[Short caption for figure 2]{\label{fig:WKS3Density} {\scriptsize Kantowski-Sachs Wigner function density projection for $\nu = 4$. The classical trajectory (near the $P_\Omega$ axis at the upper and bottom part of the Fig.) is far away from the peaks of the Wigner function.}}
\end{center}
\end{minipage}
\end{figure}

In order to extract physical information we plot the Wigner function for several values of $\nu$. We can say from Fig. \ref{FIGWKS2} and Fig. \ref{fig:WKS2Density} that the classical trajectory is near the highest peaks of the Wigner function for values close to $\nu=1$. For values of $\nu$ smaller than one we can see from Fig. \ref{fig:WKS1} and Fig. \ref{fig:WKS1Density} that there are less oscillations but the classical trajectory does not correspond to the highest peaks, in fact, there is an ample region where the Wigner function is large. For values of $\nu$ bigger than one the Fig. \ref{fig:WKS3} and Fig. \ref{fig:WKS3Density} show an increment in the number of oscillations of Wigner function and the peaks of the oscillations are far away from the classical trajectory. We can conclude that the quantum interference effects are enhanced for larger values of $\nu$.

Recently the noncommutative version of the Kantowski-Sachs model have been studied in \cite{GarciaCompean:2001wy} and it turns out interesting to analyze it using the deformation quantization formalism.
Now from the previous results we can present the noncommutative version of the Kantowski-Sachs model.  We consider a Kantowski-Sachs cosmology with a noncommutative parameter $\theta$. The operator algebra in the phase space is given by
\[
[Z_\alpha,\Pi_{Z_\beta} ] = i \hbar \delta_{\alpha \beta},
\]
\begin{equation}
[Z_\alpha,Z_\beta ] = i \theta \varepsilon_{\alpha \beta},  \ \ \ \ \ \ [\Pi_{Z_\alpha},\Pi_{Z_\beta} ] =0,
\end{equation}
where $Z_\alpha=(\Omega,\beta)$ and $\Pi_{Z_\alpha}=(P_\Omega,P_\beta)$. Of course further generalizations can be implemented in the general case with $[\Pi_{Z_\alpha},\Pi_{Z_\beta} ] \not=0.$

The noncommutative Wheeler-DeWitt equation is written as
\begin{equation}
 \left(- {\partial^2 \over \partial \Omega^2}  + {\partial^2 \over \partial \beta^2} + 48\exp(-2\sqrt{3}\Omega + \sqrt{3} \  \theta P_\beta) \right) \Psi (\Omega,\beta) = 0,
\end{equation}
where $P_\beta = -i\hbar\frac{\partial}{\partial \beta}$.
The solutions are given by \cite{GarciaCompean:2001wy}
\begin{equation}
\Psi_\nu ^{\pm} (\Omega,\beta)=e^{\pm i\nu \sqrt{3} \beta}K_{i\frac{\nu}{\hbar}} \left(\frac{4}{\hbar}\exp\bigg[-\sqrt{3}\bigg(\Omega \mp {\sqrt{3}\over 2} \nu \theta\bigg) \bigg]\right).
\end{equation}
The normalized wave function is
\begin{equation}
 \Psi^\pm (\Omega,\beta)= \frac{3^{1/4}}{\pi \hbar}\sqrt{\sinh\left(\frac{\pi \nu}{\hbar}\right)}K_{i\frac{\nu}{\hbar}} \left(\frac{4}{\hbar}\exp \bigg[{-\sqrt{3}\bigg(\Omega \mp {\sqrt{3}\over 2} \nu \theta\bigg)}\bigg]\right)
\end{equation}
which satisfy
\begin{equation}
\int_{-\infty} ^{\infty} d\Omega d \beta \Psi^*_{\nu}(\Omega, \beta)\Psi_{\nu}(\Omega, \beta)= \delta(\nu^2 -\nu '^2).
\end{equation}
Following a similar procedure as for the commutative case we are able to write the Wheeler-DeWitt-Moyal equation in the following form
\begin{align}
\nonumber
\bigg\{-\left( P_{\Omega} - \frac{i\hbar}{2}{\buildrel{\rightarrow}\over{\partial}}_\Omega\right)^2 &  + \left( P_{\beta} - \frac{i\hbar}{2}{\buildrel{\rightarrow}\over{\partial}}_\beta\right)^2  \\
& - 48 \exp{\bigg[-2\sqrt{3}\left(\Omega \mp {\sqrt{3}\over 2} \nu \theta  + \frac{i\hbar}{2}{\buildrel{\rightarrow}\over{\partial}}_{P_{\Omega}}\right)\bigg]}\bigg\} \rho^\theta (\Omega,P_{\Omega},\beta, P_{\beta}) = 0,
\end{align}
and the corresponding difference equation is
\begin{eqnarray}
\nonumber
&-&P_\Omega ^2 \rho^\theta_\Omega - \frac{12\sqrt{3}i\hbar^2 e^{-2\sqrt{3}\big(\Omega \mp {\sqrt{3}\over 2} \nu \theta \big)}}{P_\Omega}\left[\rho^\theta_\Omega(\Omega, P_\Omega + i\sqrt{3}\hbar) - \rho^\theta_\Omega(\Omega, P_\Omega - i\sqrt{3}\hbar)\right] \nonumber \\
&-&\frac{144\hbar^2 e^{-4\sqrt{3}\big(\Omega \mp {\sqrt{3}\over 2} \nu \theta \big)}}{(P_\Omega ^2 +3\hbar^2)}\left[\rho^\theta_\Omega(\Omega, P_\Omega + 2i\sqrt{3}\hbar) - 2\rho^\theta_\Omega(\Omega, P_\Omega)+ \rho^\theta_\Omega(\Omega, P_\Omega - 2i\sqrt{3}\hbar)\right] \nonumber \\
&+&\frac{144\sqrt{3}i\hbar^3 e^{-4\sqrt{3}\big(\Omega \mp {\sqrt{3}\over 2} \nu \theta \big)}}{P_\Omega (P_\Omega ^2 +3\hbar^2)}\left[\rho^\theta_\Omega(\Omega, P_\Omega + 2i\sqrt{3}\hbar) - \rho^\theta_\Omega(\Omega, P_\Omega - 2i\sqrt{3}\hbar)\right] \nonumber \\
&-& 24 e^{-2\sqrt{3}\big(\Omega \mp {\sqrt{3}\over 2} \nu \theta \big)} \left[\rho^\theta_\Omega(\Omega, P_\Omega + i\sqrt{3}\hbar) +  \rho^\theta_\Omega(\Omega, P_\Omega - i\sqrt{3}\hbar)\right] = -P_\beta ^2 \rho^\theta_\Omega.
\label{KSfinaltwo}
\end{eqnarray}
The solution to this equation is again difficult to obtain in a direct way but employing the integral representation of the Wigner function we can calculate the corresponding noncommutative Wigner function from
\footnotesize
\begin{equation}
\rho^{\theta \pm}_\Omega (\Omega, P_\Omega) = \frac{3^{1/2}}{2\pi^4 \hbar^2 }\sinh\left(\frac{\pi \nu}{\hbar}\right)\int_{-\infty} ^{\infty} dy K_{i\frac{\nu}{\hbar}} \left(\frac{4}{\hbar}e^{\pm {3 \over 2}\nu \theta}e^{-\sqrt{3}\left(\Omega -\frac{\hbar}{2}y\right)}\right) e^{-iyP_\Omega} K_{i\frac{\nu}{\hbar}} \left(\frac{4}{\hbar} e^{\pm {3 \over 2}\nu \theta}e^{-\sqrt{3}\left(\Omega +\frac{\hbar}{2}y\right)}\right).
\end{equation}
\normalsize
Thus we find
\small
\begin{align}
\nonumber
 \rho^{\theta\pm}_\Omega &(\Omega, P_\Omega) = \frac{\sinh(\pi \nu/\hbar)}{\pi^3}\frac{e^{\sqrt{3}\big(\Omega \mp {\sqrt{3}\over 2} \nu \theta \big)}}{16\hbar^{2}}\left(\frac{2}{\hbar^{2}}e^{-\sqrt{3}\big(\Omega \mp {\sqrt{3}\over 2} \nu \theta \big)}\right)^{-\frac{2iP_{\Omega}}{\sqrt{3}\hbar}}  \\
& \times G^{40}_{04}\left(\frac{16}{\hbar^{4}}e^{-4\sqrt{3} \big(\Omega \mp {\sqrt{3}\over 2} \nu \theta \big)}\bigg |  \frac{1}{4}+ i\left( \frac{\nu}{2\hbar} + \frac{P_\Omega}{\sqrt{3}\hbar}\right),\frac{1}{4} + i\left(-\frac{\nu}{2\hbar} + \frac{P_\Omega}{\sqrt{3}\hbar}\right), \frac{1}{4} + \frac{i\nu}{2\hbar}, \frac{1}{4} - \frac{i\nu}{2\hbar} \right).
\end{align}
\normalsize
Just as in the previous commutative KS case we can rewrite the former equation as
\scriptsize
\begin{align}
\nonumber
 &\rho^{\theta\pm}_\Omega (\Omega, P_\Omega) = \frac{\sinh(\pi \nu/\hbar)}{\pi^3}\frac{e^{\sqrt{3}\big(\Omega \mp {\sqrt{3}\over 2} \nu \theta \big)}}{16\hbar^{2}} \\
& \times G^{40}_{04}\left(\frac{16}{\hbar^{4}}e^{-4\sqrt{3}\big(\Omega \mp {\sqrt{3}\over 2} \nu \theta \big)}\bigg | \frac{1}{4}+ i\left( \frac{\nu}{2\hbar} + \frac{P_\Omega}{2\sqrt{3}\hbar}\right), \frac{1}{4} + i\left(\frac{-\nu}{2\hbar} + \frac{P_\Omega}{2\sqrt{3}\hbar}\right), \frac{1}{4} + i\left(\frac{\nu}{2\hbar} - \frac{P_\Omega}{2\sqrt{3}\hbar}\right), \frac{1}{4} + i\left( \frac{-i\nu}{2\hbar} - \frac{iP_\Omega}{2\sqrt{3}\hbar} \right) \right).
\end{align}
\normalsize
We can verify that this expression fulfils the equation (\ref{KSfinaltwo}) and the noncommutative effect corresponds to a displacement in the $\Omega$ variable.

\begin{figure}
\begin{minipage}[t]{8cm}
\begin{center}
\includegraphics[scale=0.65]{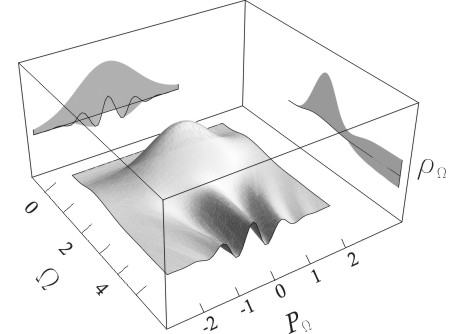}
\caption[Short caption for figure 1]{\label{fig:WCKS} {\scriptsize The Kantowski-Sachs Wigner function weighted by a Gaussian function. It can be observed the Gaussian form between $\Omega = 1$ and $\Omega = 2$ centered at $P_{\Omega} =0$ with some very smooth oscillations around $P_{\Omega}=0$.}}
\end{center}
\end{minipage}
\hfill
\begin{minipage}[t]{8cm}
\begin{center}
\includegraphics[scale=0.47]{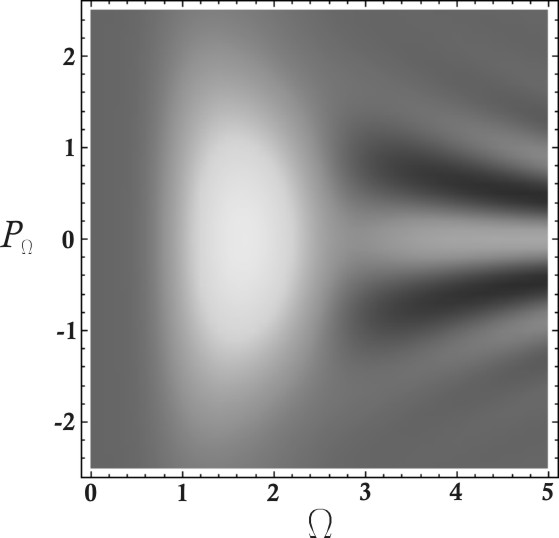}
\caption[Short caption for figure 2]{\label{fig:WCKSDensity} {\scriptsize Kantowski-Sachs Wigner function density projection weighted by a Gaussian. There is a clear maximum with some other transversal oscillations around $P_{\Omega}=0$.}}
\end{center}
\end{minipage}
\end{figure}

\begin{figure}
\begin{minipage}[t]{8cm}
\begin{center}
\includegraphics[scale=0.64]{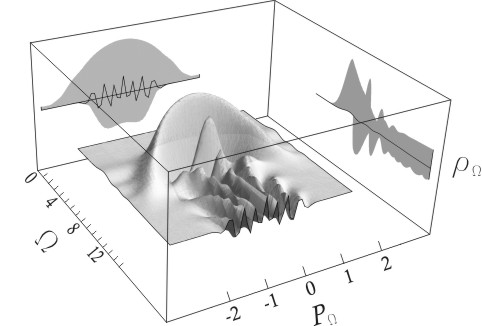}
\caption[Short caption for figure 1]{\label{FIGWNCKS} {\scriptsize Kantowski-Sachs noncommutative Wigner function weighted by a Gaussian. Several peaks of different amplitudes in the Wigner function can be observed around the Gaussian form. These are interpreted as different universes connected by tunneling.}}
\end{center}
\end{minipage}
\hfill
\begin{minipage}[t]{8cm}
\begin{center}
\includegraphics[scale=0.47]{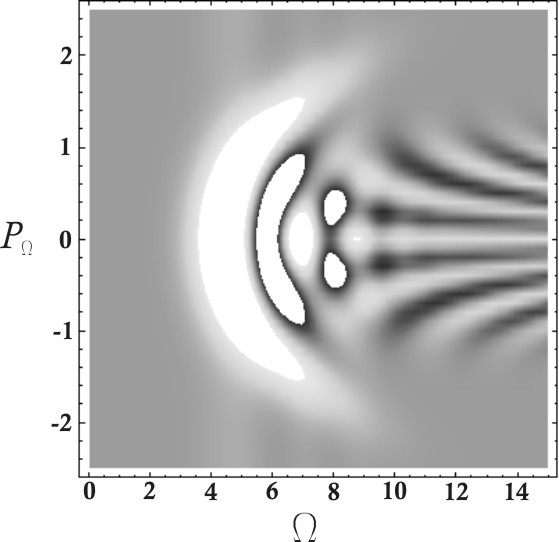}
\caption[Short caption for figure 2]{\label{fig:WNCKSDensity} {\scriptsize The density projection of the Kantowski-Sachs noncommutative Wigner function weighted by a Gaussian. Due to the noncommutativity more peaks and oscillations appears around $P_{\Omega}=0$.}}
\end{center}
\end{minipage}
\end{figure}

The noncommutativity consequences can be seen more transparently if we consider a wave packet weighted by a Gaussian
\begin{equation}
 \Psi(\Omega,\beta) = {\cal N} \int^{\infty}_{-\infty}e^{-a(\nu -b)^2}\psi^{+}_{\nu}(\Omega,\beta)d\nu.
\label{gaussNCKS}
\end{equation}

The resulting values of Eq. (\ref{gaussNCKS}) are introduced in the Wigner function (\ref{intwigner}) and the integral is performed numerically for a value of $\beta = 0$. The result is plotted in Figs. \ref{FIGWNCKS} and \ref{fig:WNCKSDensity} for the values $a=4$ and $b=1$. Figs. \ref{fig:WCKS} and \ref{fig:WCKSDensity} correspond to $\theta=0$ and Figs. \ref{FIGWNCKS} and \ref{fig:WNCKSDensity} correspond to  $\theta =4$. Here we are interested in describing which is the effect of the $\theta$ parameter. Figs.  \ref{fig:WCKS} and \ref{fig:WCKSDensity} shows that for $\theta=0$ there is only one preferred semi-classical state of the universe. Figs. \ref{FIGWNCKS} and \ref{fig:WNCKSDensity} show a significative difference consisting in having several smaller picks representing semiclassical states of lower probability.
The additional states contribute to a landscape of vacua since these new vacua can be reached by tunneling. Summarizing, at the semi-classical level, it is confirmed the observation made in Ref. \cite{GarciaCompean:2001wy} concerning that the noncommutativity of the minisuperspace leads to new possible states of the universe which contribute to its evolution at early stages.

\subsection{String cosmology with dilaton exponential potential}

Now we can treat the case of string cosmology \cite{Lidsey:1999mc} with dilaton exponential potential in the same way as in the previous subsection. We consider the $D=4$ tree level string effective action
\begin{equation}
S= \int d^4x \sqrt{-g}\left( {\cal R}_g -\frac{1}{2}(\partial \phi)^2 - V(\phi) \right),
\end{equation}
where $g_{\mu \nu}$ is the Einstein frame metric, $g$ is its determinant, ${\cal R}_g$ is the Ricci scalar compatible with $g_{\mu \nu}$, $\phi$ denotes the dilaton and we consider the dilaton potential $V(\phi) = V_0e^{\alpha \phi}$. We use the following metric
\begin{equation}
ds^2 = - \frac{N(t)^2}{a(t)^2} dt^{2} + a(t)^2 \delta_{ij} dx^i dx^j,
\end{equation}
and $N(t)$ is the lapse function. By means of the following new variables
\begin{equation}
x= \frac{1}{6}\ln\left[\frac{(u+v)(u-v)^2}{8}\right] , \hspace{2cm} z= \frac{1}{6}\ln\left[\frac{2(u+v)}{(u-v)^2}\right],
\end{equation}
where $u=\left(\frac{a^{2}}{2}\cosh 2\phi\right)$ and $v=\left(\frac{a^{2}}{2}\sinh 2\phi\right)$, the Wheeler-DeWitt equation is written in the gauge $N^{-1} = \frac{1}{2}(u-v)$ as follows \cite{Maharana}
\begin{equation}
{\cal H_{\perp}} \psi (x,z) = \left( \frac{\partial^{2}}{\partial x^2} - \frac{\partial^{2}}{\partial z^2} + 9V_0 e^{6x} \right) \psi(x,z) = 0.
\end{equation}
The solution is expressed in terms of Bessel function in the $x$ variable and as a free wave in the $z$ variable, therefore the wave function is
\begin{equation}
\psi(x,z) = e^{\pm ikz} J_{\pm ik/3}(\sqrt{V_{0}}e^{3x}),
\end{equation}
where $k$ is the separation constant. We choose the delta function normalization for the wave function. In this case it is convenient to use the wave function in the $x$ variable
\begin{equation}
\psi_{\pm k}(x)= \frac{\sqrt{3}}{2\sqrt{\sinh(\pi k)}}\left( J_{ik/3}(\sqrt{V_{0}}e^{3x}) \pm J_{-ik/3}(\sqrt{V_{0}}e^{3x})\right ),
\end{equation}
where it satisfies (see section (4.14) in \cite{Tit} and \cite{Fulling})
\begin{equation}
\int_{-\infty} ^{\infty} dx \psi^*(x)_{\pm k}\psi(x)_{\pm k'}= \delta(k^2 - k '^2).
\end{equation}

We proceed now to obtain the expression for the Wheeler-DeWitt-Moyal equation, again it is convenient to use (\ref{shift}). We get
\begin{equation}
\left[-\left( P_{x} - \frac{i\hbar}{2}{\buildrel{\rightarrow}\over{\partial}}_x\right)^2  + \left( P_{z} - \frac{i\hbar}{2}{\buildrel{\rightarrow}\over{\partial}}_z\right)^2 + 9V_0e^{\left(6\left(x + \frac{i\hbar}{2}{\buildrel{\rightarrow}\over{\partial}}_{P_{x}}\right)\right)}\right] \rho (x,P_{x},z, P_{z}) = 0,
\end{equation}
which can be separated in its real part
\begin{equation}
\left[ -P_{x}^{2} + \frac{\hbar^{2}}{4}\partial^{2}_{x} + P_{z}^{2} - \frac{\hbar^{2}}{4}\partial^{2}_{z} + 9V_0 e^{6x}\cos \left(3\hbar\partial_{P_{x}}\right)\right]\rho (x,P_{x},z, P_{z}) = 0,
\label{real2}
\end{equation}
and its imaginary part
\begin{equation}
\left[ \hbar(P_x \partial_x) - \hbar(P_z \partial_z) + 9V_0 e^{6x} \sin\left(3\hbar \partial_{P_{x}}\right)\right] \rho (x,P_{x},z, P_{z})= 0.
\label{imaginary2}
\end{equation}
If we propose $\rho = \rho_x(x,P_x) \rho_z(z,P_z)$, and consider that for a free particle in the $z$ parameter $\partial_z \rho_z = 0$, we get from equation (\ref{imaginary2})
\begin{align}
\nonumber
 \hbar P_x \partial_x ^2\rho_x(x,P_x) = & -\frac{27}{2} i V_0 e^{6x}(\rho_x (x, P_x + 3i\hbar) - \rho_x (x, P_x - 3i\hbar)) \\ \nonumber
 & -\frac{81}{4\hbar}\frac{\left(V_0 e^{6x}\right)^2}{(P_x ^2 + 9\hbar^2)}\bigg[ P_x (\rho_x(x, P_x + 6i\hbar)
 -  2\rho_x(x,P_x) + \rho_x(x,P_x -6i\hbar)) \\
  & + 3i\hbar(\rho_x(x, P_x - 6i\hbar) - \rho_x(x, P_x + 6i\hbar)) \bigg].
\end{align}
Using the former equation in (\ref{real2}) we finally obtain
\begin{align}
\nonumber
-P_z^2\rho_x(x,P_x) = & -P_x^2 \rho_x(x,P_x) + \frac{27i\hbar V_0}{4P_x} e^{6x} (\rho_x (x, P_x + 3i\hbar) - \rho_x (x, P_x - 3i\hbar)) \\ \nonumber
& + \frac{9V_0}{2} e^{6x}(\rho_x (x, P_x + 3i\hbar) + \rho_x (x, P_x - 3i\hbar)) \\ \nonumber
&-\frac{\left(9V_0e^{6x}\right)^2}{4^2P_x(P_x ^2 + 9\hbar^2)}[(P_x - 3i\hbar)\rho_x(x, P_x + 6i\hbar) \\
&- 2P_x \rho_x(x,P_x)+ (P_x + 3i\hbar)\rho_x(x, P_x - 6i\hbar)].
\label{EqndifStringCos}
\end{align}
This difference equation is complicated to solve in a direct way so as with the Kantowski-Sachs case we use the integral representation of the Wigner function. Employing the following result (see Sec. 19.3 formula (45) in \cite{ErdelyiTables}) we can calculate the Wigner function in terms of the Meijer's function
\begin{equation}
\int^\infty _0 x^{\rho -1}J_{\mu}(ax)J_{\nu}(bx^{-1})dx = 2^{\rho -1}a^{-\rho}G^{20}_{04}\left(\frac{a^{2}b^{2}}{16}\bigg | \frac{\nu}{2}, \frac{\rho + \mu}{2}, \frac{\rho - \mu}{2}, -\frac{\nu}{2}\right).
\end{equation}
Then we obtain,
\begin{eqnarray}
\rho_{x}(x, P_{x}) &=& \frac{1}{8 \hbar \pi \sin(\pi k)} \left[ G^{20}_{04}\left(\frac{V_0 ^{2} e^{12x}}{16 \hbar ^4} \bigg | \frac{i}{6}(-k+P_{x}), \frac{i}{6}(k-P_{x}), \frac{i}{6}(-k-P_{x}), \frac{i}{6}(k+P_{x}) \right)  \right. \nonumber \\
&\pm & G^{20}_{04}\left(\frac{V_0 ^{2} e^{12x}}{16 \hbar ^4} \bigg | \frac{i}{6}(-k+P_{x}), \frac{i}{6}(-k-P_{x}), \frac{i}{6}(k-P_{x}), \frac{i}{6}(k+P_{x}) \right)  \nonumber \\
&\pm &  G^{20}_{04}\left(\frac{V_0 ^{2} e^{12x}}{16 \hbar ^4} \bigg | \frac{i}{6}(k+P_{x}), \frac{i}{6}(k-P_{x}), \frac{i}{6}(-k-P_{x}), \frac{i}{6}(-k+P_{x}) \right)  \nonumber \\
&+& \left. G^{20}_{04}\left(\frac{V_0 ^{2} e^{12x}}{16 \hbar ^4} \bigg | \frac{i}{6}(k+P_{x}), \frac{ik}{6}(-k-P_{x}), \frac{i}{6}(k-P_{x}), \frac{i}{6}(-k+P_{x}) \right) \right].
\label{WKSMeijer}
\end{eqnarray}
We want to note, that it is also possible to write the Wigner function in terms of the hypergeometric function
${}_0 F_{3}$ employing \cite{Grad}. This is an straightforward but long calculation that we will not present here.

It can be verified that $\rho_x$ given by Eq. (\ref{WKSMeijer}) satisfies the equation (\ref{EqndifStringCos}).

\subsection{Baby Universes}

In this subsection we will consider another example of the use of the
deformation quantization formalism to wormhole solutions in general
relativity \cite{Hawking:1988ae,Hawking:1990jb,Lyons:1991im}.
Here we will have a system with two coordinates of the
{\it flat} minisuperspace (2 degrees of freedom).

We are going to consider a baby universe with conformal matter
$\phi_0$ and a three-metric $h_{ij}$ defined on a Cauchy
hypersurface ${\cal S}$ of a closed wormhole universe. The matter is
represented by a conformal invariant scalar field expanded in
hyper-spherical harmonics $Q_n$ of the surface ${\cal S}$ given by
$\phi_0= \frac{1}{a} \sum_n f_n Q_n$, where $a$ is the scale factor and
$f_n$ are orthonormal modes. The metric on ${\cal S}$ is given by
$h_{ij}= a^2 \cdot (\Omega_{ij} + \varepsilon_{ij}).$ Here
$\Omega_{ij}$ is the metric of a unit three-sphere ${\bf S}^3$ and
$\varepsilon_{ij} = \sum_n \big( a_n \Omega_{ij}Q_n + b_n P_{ijn} +
c_{n} S_{ijn} + d_{n} G_{ijn} \big)$. The $Q_n$'s are the scalar harmonics of the 3-sphere, $P_{ijn}$
is a suitable combination of $Q_n$, $S_{ijn}$ is defined in terms of the transverse
vector harmonics and $G_{ijn}$ are the transverse traceless tensor harmonics on ${\cal S}$.

On the gravitational part, in
a suitable gauge ($a_n=b_n=c_n=0$) of $h_{ij}$ on ${\cal S}$ and
considering the case without gravitons ($d_n=0$), one can express
the three metric simply as: $h_{ij}= a^2 \cdot \Omega_{ij}$. Thus,
the wave function $\Psi$ is a function of the scale factor $a$ and
the harmonic modes of the scalar field $f_n$. This wave function
fulfills the Wheeler-DeWitt equation (\ref{WdWeqnm}) of the form

\begin{equation}
\bigg[\sum_n \bigg( -\frac{\partial^2}{\partial f_n^2} + n^2 f_n^2  \bigg) -
\bigg(-\frac{\partial^2}{\partial a^2} + a^2  \bigg) \bigg] \Psi(a,f_n) = 0,
\end{equation}
where we have implemented the canonical relation on the momenta $\widehat{p}_{f_n} =
-i \hbar \frac{\partial}{\partial f_n}$ and $\widehat{p}_{a} = -i
\hbar \frac{\partial}{\partial a}$.

In the context of quantum cosmology the solution factorizes in a
purely gravitational part and an purely matter part. Both of them
correspond to harmonic oscillators and the solution is given by
\begin{equation}
\Psi(a,f_n) = A H_m(a)\exp\left(- \frac{a^2}{2 }\right)  \cdot
\prod_{n} H_{m_n}(f_n \sqrt{n}) \exp\left(- \frac{n f^2_n}{2
}\right),
\end{equation}
where $H_m(x)$ are the Hermite polynomials and $A$ is a
normalization constant. This solution satisfies the boundary
conditions such that: $\psi(a,f_n) \to 0$ as $a \to \infty$ and it is
regular at $a=0$.

Here it is assumed, as in \cite{Hawking:1990jb}, that the zero-point
energy of the gravitational sector will
precisely compensates the zero-point energy of the matter
oscillators as it happens in a supersymmetric theory. This solution
represents a closed universe carrying $m$ scalar particles in the
$n$-th mode. Thus the ground state $|\Psi_0 \rangle$ will correspond
to $m=0$ and $n=0$, i.e., the absence of matter particles
and consequently excited states of the scale factor part.

The wave function factorization $\Psi(a,f_n) = \psi_0(a) \cdot
\prod_n \psi_n(f_n)$ comes from the inner product between $\langle a,
f_n |$ and $|\psi_0, \psi_n \rangle$. The ground state is given by
$|\Psi_0 \rangle = |\psi_0 \rangle_a \otimes  |\psi_0\rangle_f$.

Let $\widehat{b}$ and $\widehat{b}^\dag$ the annihilation and
creation operators of the $m_a$-th modes of the gravitational sector
defined by: $\widehat{b} |0_a,m_f\rangle =0$ and
$[\widehat{b}^\dag]^{m_a} |\Psi_0\rangle = |m_a,m_f\rangle$. Now,
let $\widehat{d}$ and $\widehat{d}^\dag$ the annihilation and
creation operators of the $m_f$-th modes of scalar particles defined
by: $\widehat{d} |m_a,0_f\rangle =0$ and $[\widehat{d}^\dag]^{m_f}
|\Psi_0\rangle = |m_a,m_f\rangle$. The combination yields
$\widehat{b} \cdot \widehat{d} |0_a,0_f\rangle =0$ and
$[\widehat{b}^\dag]^{m_a} \cdot
[\widehat{d}^\dag]^{m_f}|\Psi_0\rangle = |m_a,m_f\rangle$.

In the WWGM formalism we have the following
general stationary Wheeler-DeWitt-Moyal equation
\begin{equation}
H_B \star \rho_{W}(a,p_a,f_n,p_{f_n}) = 0,
\label{BUstar}
\end{equation}
where $\rho_W =\rho_W(a,p_a,f_n,p_{f_n})$ is the Wigner function and
$H_B$ stands for the baby universe Hamiltonian
\begin{equation}
H_B = H_f + H_a,
\end{equation}
where
\begin{equation}
H_f= \sum_n \big( p^2_{f_n} + \omega^2_n f^2_n \big)
\end{equation}
with $\omega_n =n$ and
\begin{equation}
H_a = p^2_a + a^2.
\end{equation}
The Moyal product is given by
\begin{equation}
f\star g= f\exp \left( {i \hbar \over
2}\stackrel{\leftrightarrow}{\cal P}_B\right) g,
\end{equation}
where the corresponding Poisson operator has the following form
\begin{equation}
\stackrel{\leftrightarrow}{\cal P}_B = \stackrel{\leftrightarrow}{\cal P}_f +
 \stackrel{\leftrightarrow}{\cal P}_a
 =\sum_n \bigg(\frac{\stackrel{\leftarrow}\partial}{\partial f_n}
 \frac{\stackrel{\rightarrow}\partial}{\partial p_{f_n}} -
 \frac{\stackrel{\leftarrow}\partial}{\partial p_{f_n}}
 \frac{\stackrel{\rightarrow}\partial}{\partial f_n} \bigg) + \frac{\stackrel{\leftarrow}\partial}{\partial a}  \frac{\stackrel{\rightarrow}\partial}{\partial p_a} -
\frac{\stackrel{\leftarrow}\partial}{\partial p_a}  \frac{\stackrel{\rightarrow}\partial}{\partial a}.
\end{equation}
We can write down $\rho_{W0}= \rho^a_{W0} \cdot \rho^f_{W0}$ then we
can separate (\ref{BUstar}) into two parts
\begin{equation}
H_a \star \rho^a_{W0}(a,p_a) = -E \rho^a_{W0}(a,p_a),
\label{a}
\end{equation}
and
\begin{equation}
H_f \star \rho^f_{W0}(f,p_f) = E\rho^f_{W0}(f,p_f).
\label{b}
\end{equation}
Therefore we have that the equation (\ref{BUstar}) at the order $\hbar$
can be written as
\begin{equation}
\sum_n \bigg(\omega_n^2 f_n \cdot \frac{\partial
\rho^f_{W0}}{\partial p_{f_n}} - p_{f_n} \cdot \frac{\partial
\rho^f_{W0}}{\partial f_n}\bigg) - a \cdot \frac{\partial
\rho^a_{W0}}{{\partial p_a}} + p_a \cdot \frac{\partial
\rho^a_{W0}}{\partial a} =0,
\end{equation}
where $\rho_{W0}$ is the Wigner function for the ground state. Thus
the solution to this equation is given by
\small
\begin{equation}
\rho_{W0}(p_a,a,p_{f_n},f_n) = \rho^a_{W0}(a,p_a) \cdot
\rho^f_{W0}(f_n, p_{f_n}) = A \exp\left[-{2 \over \hbar}\big(p_a^2 +
a^2 \big)\right] \cdot \prod_n \exp \bigg[-{2 \over \hbar}\big(
p^2_{f_n} + \omega_n^2 f_n^2\big) \bigg].
\end{equation}
\normalsize
The density matrix for all excited states is given by
\begin{equation}
\widehat{\rho}_{r,s} = [\widehat{b}^\dag]^r \cdot
[\widehat{d}^\dag]^s|\Psi_0\rangle \langle \Psi_0 |
 [\widehat{d}^\dag]^s \cdot [\widehat{b}^\dag]^r.
\end{equation}
The WWGM formalism allows to compute from this density matrix the
general Wigner function for all excited states \cite{campos}
\begin{equation}
[\rho_W]_{m,n} =  [{b}^*] \star  \cdots  \star [{b}^*] \star  [{d}^*] \star  \cdots \star [{d}^*]  \star  \rho_{W0} \star [{d}^*] \star \cdots \star [{d}^*] \star [{b}^*] \star \cdots \star [{b}^*].
\end{equation}
It is straightforward to show that it leads to the solution
\footnotesize
\begin{equation}
\rho_{W}(p_a,a,f_n,p_{f_n}) =  A  \exp\left[-{2 \over
\hbar}\big(p_a^2 + a^2 \big)\right] L_m\left( {4\over \hbar}(p_a^2 +
a^2) \right) \cdot \prod_n
 \exp \bigg[-{2 \over \hbar}\big(
p^2_{f_n} + \omega_n^2 f_n^2\big) \bigg] L_{m_n}\left({4 \over \hbar}
(p^2_{f_n} + \omega_n^2 f_n^2) \right).
\end{equation}
\normalsize
Here $L_m(x)$ is the Laguerre polynomial of degree $m$. Remember
that they are related to the Hermite polynomials through the
familiar formula $L_n(x^2 + y^2) = (-1)^n 2^{-2n} \sum_{m=0}^n
\frac{1}{m! (n-m)!} H_{2m}(x) H_{2n-2m}(y).$ The case for the
minimal coupling scalar matter follows a similar treatment and will
not be discussed here, but analogous formulas can be obtained for
this case.


\section{Final Remarks}

In this paper we have constructed the WWGM formalism in the flat
superspace (and phase superspace). The WWGM correspondence is explicitly developed and the
Stratonovich-Weyl quantizer, the star product and the Wigner
functional are obtained. These results can be used in general situations but in a first approach we applied the formalism to some interesting minisuperspace models widely studied in the literature, in particular, we used the Moyal star product to describe some relevant cosmological models in phase space.

We studied de Sitter quantum cosmology using the Hartle-Hawking, Vilenkin and Linde boundary conditions, where we have found numerically the Wigner function for all of these cases.

For the Hartle-Hawking wave function the behavior of the Wigner function  presents many oscillations due to interference terms between the wave functions of expanding and contracting universes. Similarly as in Ref. \cite{Habib:1990hx} our result shows that the highest peaks of the Wigner function do not coincide with the classical trajectory of the universe.

The Linde Wigner function has a similar behavior to the Hartle-Hawking case, where there are also expanding and contracting components of the wave function. The main difference in their corresponding Wigner functions is just a sign in the interference terms between the expanding and contracting wave functions and as a consequence it produces a reduction in the amplitude of the oscillations inside the classical region.

In the case of the Vilenkin tunneling wave function we notice that the Wigner function has less oscillations compared to the Hartle-Hawking Wigner function. This is explained by the fact that there are less oscillation effects because there is only an outgoing wave.  In this case the classical trajectory corresponds almost to the maxima of the peaks of the Wigner function.

However the classical limit for these three models is difficult to obtain due to the existence of oscillations in the Wigner functions. It is important to note that decoherence of the Vilenkin Wigner function is in principle simpler to obtain
since the interference terms are absent because there is only an expanding universe around $p=0$.

For the Kantowski-Sachs cosmological model we found the Wheeler-DeWitt-Moyal equation which is equivalent to a differential-difference equation. We found its exact solution in terms of the Meijer's function.   We observe that the classical trajectory corresponds to the highest peaks of the Wigner function for values close to $\nu=1$. The situation for $\nu \leq 1$ in the Wigner function corresponds to have less oscillations and the classical trajectory does not correspond to the highest peaks. For values of $\nu \geq 1$ we have an increment in the number of oscillations of the Wigner function and we do not have the peaks of the oscillations near the classical trajectory. Thus it seems that $\nu$ could be regarded as a parameter encoding the quantum interference effects.

We have also considered the noncommutative Kantowski-Sachs model. In a similar way we obtain the analytic Wigner function and its differential-difference equation. This case
presents a noncommutative parameter $\theta$ deforming the Wheeler-DeWitt-Moyal equation. The Wigner function is determined in terms of the above Meijer's function with shifted argument by a factor proportional to $\theta$. We have constructed numerically the Wigner function with wave packet weighted by a Gaussian to see the effects of the noncommutativity. There were found several peaks in the Wigner function around the
Gaussian which can be interpreted as different universes connected by tunneling. Thus, at the semi-classical level, the statement made in \cite{GarciaCompean:2001wy} about that the noncommutativity of the minisuperspace produces new possible states of the universe at early stages is confirmed.

String cosmology with dilaton exponential potential is also discussed. We found the corresponding Wheeler-DeWitt-Moyal equation and the equivalent differential-difference equation. These equations have an exact solution in terms of the hypergeometric and Meijer's functions.

Baby universe solutions are also obtained in this context where the Wigner function is calculated by finding the exact solution of its Wheeler-DeWitt-Moyal equation consisting in two decoupled deformed harmonic oscillators in terms of Laguerre polynomials.

It is important to remark that this work opens the possibility of
treating several important questions that remain unsolved in quantum
cosmology with a novel approach. For instance, deformation quantization allows to deal with systems having phase spaces with nontrivial topology. In quantum cosmology, the
existence of symmetries implies that the phase-superspace and in
particular, the phase-minisuperspace will be reduced by the implementation of these symmetries leading to a  non-trivial topological space with a non-flat metric. Therefore, deformation quantization is able to treat these mentioned cases in a natural way. Moreover the extension to supersymmetric quantum cosmology \cite{Macias:1993mq,Obregon:1996dt,D'Eath:1996at} can be also treated applying the results of \cite{fermi}.

Another point to remark is that the Wheeler-DeWitt-Moyal equation (\ref{QHconstraint}) proposed in the present paper contains the generalized mass-shell equation and the Wheeler-DeWitt-Vlasov transport equation in the flat minisuperspace  \cite{Calzetta:1989vk} encoded in its
real an imaginary parts respectively. The former result was obtained using the Moyal $\star-$product and then all the technics of deformation quantization developed for a long time can be apply to it.  It is known that the mass-shell equation and the  Wheeler-DeWitt-Vlasov transport equation admits a suitable generalization to curved minisuperspace \cite{Calzetta:1989vk}. It would be interesting to deal with curved symplectic cases where the Fedosov's approach could be applied in order to find the Wheeler-DeWitt-Moyal equation in non-trivial phase superspaces and obtain the mass-shell and the  Wheeler-DeWitt-Vlasov equations. We will study this problem in a further communication.

Besides, as was mentioned before, the problem of obtaining the classical limit by implementing a coarse graining is relevant. It is possible to model a coarse graining through the Liouville equation with friction and diffusion terms \cite{Habib:1990hx}, this approach can be addressed in the context of the deformation quantization formalism.

To conclude, we consider that deformation quantization possesses
various advantages in order to deal with more complicated problems in quantum cosmology for example, to treat systems with non trivial topology or with curved minisuperspaces. For these cases the canonical quantization could lead to the existence of nonhermitian operators which is avoided in deformation quantization as a result of the use of classical objects instead of operators. For these reasons more examples and further research
is needed to develop this approximation.

\vskip 1truecm 
\centerline{\bf Acknowledgments}

\vskip 1truecm

The work of R. C., H. G.-C. and F. J. T. was partially supported by SNI-M\'exico, CONACyT research grants: J1-60621-I, 103478 and 128761. In addition R. C. and F. J. T. were partially supported by COFAA-IPN and by SIP-IPN grants 20100684 and 20100887. We are indebted to H\'ector Uriarte for all his help in the elaboration of the figures presented in the paper.




\end{document}